\documentclass[aps,prc,superscriptaddress,floats,floatfix,reprint,reprintnumbers]{revtex4}
\usepackage{blindtext}
\pdfoutput=1
\usepackage{epsfig}
\usepackage{graphicx}
\usepackage{dcolumn}
\usepackage{bm}
\usepackage{booktabs}
\usepackage{color}
\usepackage{mathrsfs}
\usepackage{ulem}
\usepackage{epstopdf}
\graphicspath{{plots/}}
\usepackage{verbatim}
\usepackage{indentfirst}
\usepackage{float}
\usepackage[colorlinks,
           bookmarks=true,
           linkcolor=blue,
           urlcolor=blue,
            anchorcolor=black,
            citecolor=blue]{hyperref}
\usepackage[bf,raggedright,indentafter]{titlesec}
\usepackage{rotating}
\usepackage{booktabs}
\usepackage{tabularx}
\usepackage{makecell}
\usepackage{hypcap}
\usepackage{subfigure}%

\renewcommand{\baselinestretch}{1.47}

\begin{document}
\title{\Large Testing  SU(3) Flavor Symmetry in Semileptonic  and Two-body Nonleptonic  Decays of Hyperons}
\author{Ru-Min Wang$^{1,\dagger}$,~~Mao-Zhi Yang$^{2,*}$,~~Hai-Bo Li$^{3,4,\S}$~  and~ Xiao-Dong Cheng$^{5,\ddag}$\\
$^1${\scriptsize College of Physics and Communication Electronics, JiangXi Normal University, NanChang, JiangXi 330022, China}\\
$^2${\scriptsize School of Physics, Nankai University, TianJin 300071, China}\\
$^3${\scriptsize Institute of High Energy Physics, BeiJing 100049,  China}\\
$^4${\scriptsize University of Chinese Academy of Sciences, BeiJing 100049, China}\\
$^5${\scriptsize College of Physics and Electronic Engineering, XinYang Normal University, XinYang, Henan 464000, China}\\
$^\dagger${\scriptsize ruminwang@sina.com}\\
$^*${\scriptsize yangmz@nankai.edu.cn}\\
$^\S${\scriptsize lihb@ihep.ac.cn}\\
$^\ddag${\scriptsize chengxd@mails.ccnu.edu.cn}\\
}

\begin{abstract}
The semileptonic decays and two-body nonleptonic decays of light baryon octet ($T_8$) and decuplet ($T_{10}$)  consisting of light $u,d,s$ quarks are studied with the SU(3) flavor symmetry in this work. We obtain  the amplitude relations  between different decay modes by the SU(3)  irreducible representation approach, and then predict relevant branching ratios by present experimental data within $1 \sigma$ error.
We find that the predictions for all branching ratios except $\mathcal{B}(\Xi\rightarrow \Lambda^0\pi)$ and $\mathcal{B}(\Xi^*\rightarrow \Xi\pi)$ are in good agreement with present experimental data,  that implies
 the neglected $C_+$ terms or SU(3) breaking effects might contribute at the order of a few percent in  $\Xi\rightarrow \Lambda^0\pi$ and $\Xi^*\rightarrow \Xi\pi$ weak decays.
We predict that $\mathcal{B}(\Xi^{-}\rightarrow \Sigma^0\mu^-\bar{\nu}_\mu)=(1.13\pm0.08)\times10^{-6}$, $\mathcal{B}(\Xi^{-}\rightarrow \Lambda^0\mu^-\bar{\nu}_\mu)=(1.58\pm0.04)\times10^{-4}$, $\mathcal{B}(\Omega^-\rightarrow\Xi^0\mu^-\bar{\nu}_\mu)=(3.7\pm1.8)\times10^{-3}$, $\mathcal{B}(\Sigma^-\rightarrow \Sigma^0e^-\bar{\nu}_e)=(1.35\pm0.28)\times10^{-10}$,  $\mathcal{B}(\Xi^-\rightarrow \Xi^0e^-\bar{\nu}_e)=(4.2\pm2.4)\times10^{-10}$. We also study  $T_{10}\to T_8 P_8$ weak, electromagnetic or strong decays. Some of these decay modes could be observed by the BESIII, LHCb and other experiments in the near future.  Due to the very small life times of $\Sigma^0$,  $\Xi^{*0,-}$, $\Sigma^{*0,-}$ and $\Delta^{0,-}$,  the branching ratios  of these baryon weak decays are only at the order of $\mathcal{O}(10^{-20}-10^{-13}$), which are too small to be reached  by current experiments. Furthermore, the longitudinal branching ratios of  $T_{8A} \to T_{8B} \ell^- \bar{\nu}_\ell~(\ell=\mu,e)$  decays are also given.
\end{abstract}
\maketitle

\section{INTRODUCTION}
A lot of  semileptonic decays and two-body nonleptonic decays  of  light octet baryons (such as $\Xi^-\rightarrow\Sigma^0e^-\bar{\nu}_e$, $\Xi^-\rightarrow\Lambda^0\ell^-\bar{\nu}_\ell$, $\Xi^0\rightarrow\Sigma^+\ell^-\bar{\nu}_\ell$, $\Lambda^0\rightarrow p\ell^-\bar{\nu}_\ell$, $\Sigma^-\rightarrow n\ell^-\bar{\nu}_\ell$, $\Sigma^-\rightarrow \Lambda^0e^-\bar{\nu}_e$, $\Sigma^+\rightarrow \Lambda^0e^+\nu_e$, $n\rightarrow pe^-\bar{\nu}_e$, $\Sigma^+\rightarrow p\pi^0$,
$\Sigma^+\rightarrow n\pi^+$, $\Sigma^-\rightarrow n\pi^-$, $\Lambda^0\rightarrow p\pi^-$, $\Lambda^0\rightarrow n\pi^0$, $\Xi^-\rightarrow \Lambda^0\pi^-$, $\Xi^0\rightarrow \Lambda^0\pi^0$) and a few light decuplet baryon decays  (such as $\Omega^-\rightarrow\Xi^0e^-\bar{\nu}_e,\Xi^0\pi^-,\Xi^-\pi^0,\Lambda^0K^-$) were measured  a long time ago by SPEC, HBC, OSPK etc \cite{PDG2018}. Now the sensitivity for measurements of $\Lambda,\Sigma,\Xi,\Omega$ hyperon decays  is in the range of $10^{-5}-10^{-8}$ at the BESIII \cite{Li:2016tlt,Bigi:2017eni,Asner:2008nq,BESIIILi2019}, and these hyperons are also produced copiously  at the LHCb experiment~\cite{Aaij:2017ddf,Junior:2018odx}. Besides confirming information
obtained earlier  by SPEC, HBC, OSPK etc., new information on light baryon decays will be obtained at the BESIII  and LHCb experiments.
The precise measurements of these decays are of great importance  in   determining the V-A structure and quark-flavor mixing of charged current weak interactions \cite{Weinberg:2009zz,Severijns:2006dr,Cabibbo:1963yz} as well as probing  the non-standard charged current interactions \cite{Cirigliano:2012ab,Chang:2014iba}.

Theoretically, the factorization does not work well for  $s,d$ quark decays   since $s,d$ quarks are very light and can not use the heavy quark expansion. There is no reliable method to calculate these decay matrix elements  at present. In the lack of reliable calculations, the  symmetry analysis can provide very useful
information about the decays. SU(3) flavor symmetry is one of the symmetries
which have attracted a lot of attentions. The SU(3) flavor  symmetry approach, which is independent of the detailed dynamics,  offers an opportunity to relate different decay modes.  Nevertheless, it cannot determine
the size of the amplitudes by itself. However, if experimental data are enough, one may use the data to extract the SU(3) irreducible amplitudes, which can be viewed as predictions based on symmetry.
There are two popular ways of the SU(3) flavor symmetry. One is to construct the
SU(3) irreducible representation amplitude by decomposing effective Hamiltonian. The other way is topological diagram approach, where decay amplitudes are represented by connecting
quark line flows in different ways  and then relate them by the SU(3) symmetry.

The SU(3) flavor symmetry  works well in heavy hadron decays, for instance,  the $b$-hadron decays  \cite{He:1998rq,He:2000ys,Fu:2003fy,Hsiao:2015iiu,He:2015fwa,He:2015fsa,Deshpande:1994ii,Gronau:1994rj,Gronau:1995hm,Shivashankara:2015cta,Zhou:2016jkv,Cheng:2014rfa}   and   the $c$-hadron decays  \cite{Grossman:2012ry,Pirtskhalava:2011va,Cheng:2012xb,Savage:1989qr,Savage:1991wu,Altarelli:1975ye,Lu:2016ogy,Geng:2017esc,Geng:2018plk,Geng:2017mxn,Geng:2019bfz,Wang:2017azm,Wang:2019dls,Wang:2017gxe,Muller:2015lua}.
The experimental data of some semileptonic hyperon decays are well explained by the Cabibbo theory \cite{Cabibbo:1963yz}, which assumes the SU(3) symmetry breaking effects are neglected.
The SU(3) flavor symmetry breaking effects are also studied in the hyperon beta-decays \cite{Gaillard:1984ny,Carrillo-Serrano:2014zta,Pham:2012db,FloresMendieta:1998ii}, where it is found that the SU(3) symmetry breaking effects in these decays are small.
In this paper, we will systematically study  $T_{8,10}\to T_{8}\ell^-\bar{\nu}_\ell$  and $T_{8,10}\to T_{8} P$ decays  by the SU(3) irreducible representation approach (IRA).    We  will  firstly construct the
SU(3) irreducible representation amplitudes for different kinds of $T_{8}$ and $T_{10}$ decays,  secondly  obtain the decay amplitude relations between different decay modes, then use the available data to extract the SU(3) irreducible amplitudes,  and finally  predict the  not-yet-measured modes for further tests in experiments.

This paper is organized as follows. In Sec. II, the semileptonic weak decays of  the $T_{8,10}$ hyperons are studied. In Sec. III, we will explore the two-body nonleptonic decays of hyperons which are through weak interaction, electromagnetic or strong interaction.  Our conclusions are given in Sec. IV.

\section{Semileptonic decays of hyperons}

The light baryons $T_{8}$ ($T_{10}$), which are
octet (decuplet)  under the SU(3) flavor symmetry of $u,d,s$ quarks,  can be written as
\begin{eqnarray}
 T_8&=&\left(\begin{array}{ccc}
\frac{\Lambda^0}{\sqrt{6}}+\frac{\Sigma^0}{\sqrt{2}} & \Sigma^+ & p \\
\Sigma^- &\frac{\Lambda^0}{\sqrt{6}}-\frac{\Sigma^0}{\sqrt{2}}  & n \\
\Xi^- & \Xi^0 &-\frac{2\Lambda^0}{\sqrt{6}}
\end{array}\right)\,,\\
 T_{10}&=&\frac{1}{\sqrt{3}}\left(
 \left(\begin{array}{ccc}
\sqrt{3}\Delta^{++} & \Delta^{+}  & \Sigma^{*+}  \\
\Delta^{+} & \Delta^{0}  & \frac{\Sigma^{*0}}{\sqrt{2}}  \\
\Sigma^{*+} & \frac{\Sigma^{*0}}{\sqrt{2}}  &\Xi^{*0}
\end{array}\right),
\left(\begin{array}{ccc}
\Delta^{+} & \Delta^{0}  & \frac{\Sigma^{*0}}{\sqrt{2}}  \\
\Delta^{0} & \sqrt{3}\Delta^{-}  & \Sigma^{*-} \\
\frac{\Sigma^{*0}}{\sqrt{2}} & \Sigma^{*-} &\Xi^{*-}
\end{array}\right),
\left(\begin{array}{ccc}
\Sigma^{*+} & \frac{\Sigma^{*0}}{\sqrt{2}}  &\Xi^{*0}  \\
\frac{\Sigma^{*0}}{\sqrt{2}} & \Sigma^*{-}  & \Xi^{*-} \\
\Xi^{*-} & \Xi^{*-} &\sqrt{3}\Omega^{-}
\end{array}\right)\right).
\end{eqnarray}
In this section, we focus on $\Delta S=0$ and $\Delta S=1$ semileptonic  decays  of  hyperons,  which decay through $d\to u e^-\bar{\nu}_e$ or $s\to u \ell^-\bar{\nu}_\ell$ transitions, respectively. Since $\Delta S=2$ semileptonic decays are forbidden, we will not study them in this work.

\subsection{$T_{8A}\to T_{8B}\ell^-\bar{\nu}_\ell$ semileptonic decays}
\label{S88}
\begin{figure}[b]
\centering
\includegraphics[scale=0.9]{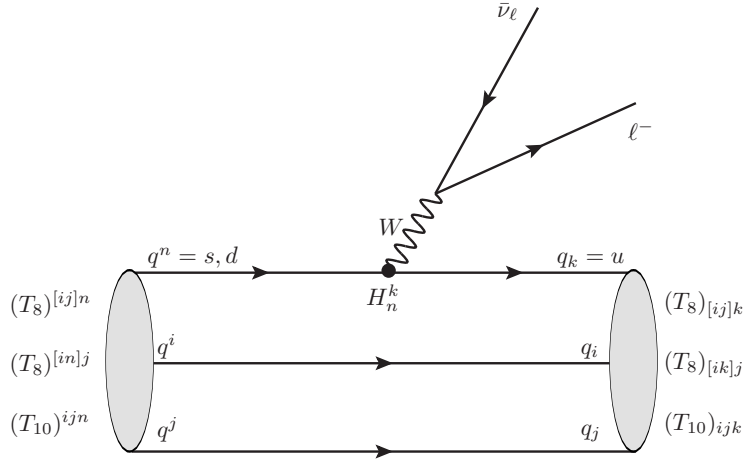}
\caption{Feynman diagram for semileptonic weak decays  $T_{8,10}\to T_{8}\ell\bar{\nu}_\ell$. }
\label{fig:SL}
\end{figure}

In the Standard Model (SM), the feynman diagram for $T_{8A}\to T_{8B}\ell^-\bar{\nu}_\ell$ decays is shown in  Fig. \ref{fig:SL}, and the amplitudes of $T_{8A}\to T_{8B}\ell^-\bar{\nu}_\ell$  can be written as \cite{Kadeer:2005aq}
\begin{eqnarray}
{\cal A}(T_{8A}\to T_{8B} \ell^- \bar{\nu}_\ell)&=&\sum_{\lambda_{W},\lambda'_{W}=\pm 1,0 ,t}\frac{G_F}{2\sqrt{2}} H_{\lambda_B \lambda_{W}} \bar{u}_{\ell}\gamma_{\beta}(1-\gamma_5)v_\nu \epsilon^{*\beta}(\lambda'_{W})g_{\lambda_{W}\lambda'_{W}},
\end{eqnarray}
with
\begin{eqnarray}
H_{\lambda_B \lambda_{W}}&=&H^{V}_{\lambda_B \lambda_{W}}-H^{A}_{\lambda_B \lambda_{W}}, ~~~~H^{V(A)}_{\lambda_B \lambda_{W}}=\langle  T_{8B}|J_{\mu }^{V(A)}|T_{8A}\rangle \epsilon^{\mu}(\lambda_{W}),\nonumber\\
H^V_{\frac{1}{2}0}&=&\frac{\sqrt{Q_-}}{\sqrt{q^2}}\Big[(m_{A}+m_{B})f_1({q^2})-q^2f_2(q^2)\Big],\nonumber\\
H^A_{\frac{1}{2}0}&=&\frac{\sqrt{Q_+}}{\sqrt{q^2}}\Big[(m_{A}-m_{B})g_1({q^2})+q^2g_2(q^2)\Big],\nonumber\\
H^V_{\frac{1}{2}1}&=&\sqrt{2Q_-}\Big[-f_1({q^2})+(m_{A}+m_{B})f_2(q^2)\Big],\nonumber\\
H^A_{\frac{1}{2}1}&=&\sqrt{2Q_+}\Big[-g_1({q^2})-(m_{A}-m_{B})g_2(q^2)\Big],\nonumber\\
H^V_{\frac{1}{2}t}&=&\frac{\sqrt{Q_+}}{\sqrt{q^2}}\Big[(m_{A}-m_{B})f_1({q^2})+q^2f_3(q^2)\Big],\nonumber\\
H^A_{\frac{1}{2}t}&=&\frac{\sqrt{Q_-}}{\sqrt{q^2}}\Big[(m_{A}+m_{B})g_1({q^2})-q^2g_3(q^2)\Big],\label{Eq:HVA}
\end{eqnarray}
where $q=p_A-p_B$ and $Q_\pm =(m_{A} \pm m_{B})^2-q^2$.  Either from parity or from explicit calculation, we have the relations $H^V_{-\lambda_2 -\lambda_1}=H^V_{\lambda_2 \lambda_1}$, $H^A_{-\lambda_2 -\lambda_1}=-H^A_{\lambda_2 \lambda_1}$. The form factors $f_i(q^2)$ and $g_i(q^2)$ are defined by  \cite{Shivashankara:2015cta}
\begin{eqnarray}
<T_{8B}(p_B,\lambda_B)|\bar{c}\gamma_\mu b|T_{8A}(p_A,\lambda_A)>&=&\bar u_B(p_B,\lambda_B)[f_1(q^2)\gamma_{\mu}+if_2(q^2){\sigma}_{\mu\nu}q^{\nu}+f_3(q^2)q_{\mu}]u_A(p_A,\lambda_A),
\label{Lb2Lc1}\nonumber\\
<T_{8B}(p_B,\lambda_B)|\bar{c}\gamma_\mu \gamma^5 b|T_{8A}(p_A,\lambda_A)>&=&\bar u_B(p_B,\lambda_B)[g_1(q^2)\gamma_{\mu}+ig_2(q^2){\sigma}_{\mu\nu}q^{\nu}+g_3(q^2)q_{\mu}]\gamma_{5}u_A(p_A,\lambda_A).
\end{eqnarray}

In term of the SU(3) IRA, the helicity amplitudes $H^{V(A)}_{\lambda_B \lambda_{W}}$ can be parameterized as
\begin{eqnarray}
H^{V(A)}_{\lambda_B \lambda_{W}}&=&~~a_{n1}H^k_n(T_8)^{[ij]n}(T_8)_{[ij]k}+a_{n2}H^k_n(T_8)^{[ij]n}(T_8)_{[ik]j}+a_{n3}H^k_n(T_8)^{[in]j}(T_8)_{[ij]k}\nonumber\\
                 &&+a_{n4}H^k_n(T_8)^{[in]j}(T_8)_{[ik]j}+a_{n5}H^k_n(T_8)^{[in]j}(T_8)_{[kj]i}, \label{EQ:HVAIRA}
\end{eqnarray}
where $H^k_n=V_{q_kq_n}$ is the CKM matrix element, $a_{ni}\equiv(a_{ni})^{V(A)}_{\lambda_B\lambda_{W}}(q^2)$ are  the nonperturbative coefficients,  and $n=2(3)$ for $q^n=d(s)$.
The SU(3) IRA helicity amplitudes $H^{V(A)}_{\lambda_B \lambda_{W}}$ are listed in the second  column of Table \ref{Tab:ASLT8}.
\begin{table}[b]
\renewcommand\arraystretch{1.5}
\tabcolsep 0.25in
\centering
\caption{The helicity amplitudes $H^{V(A)}_{\lambda_B \lambda_{W}}$ of $T_{8A}\to T_{8B}\ell^-\bar{\nu}_\ell$ decays.}\vspace{0.1cm}
{\footnotesize
\begin{tabular}{rcc}  \hline\hline
$H^{V(A)}_{\lambda_B\lambda_W}$~~~~~~~~~~~~& SU(3) IRA amplitudes & Reparameterization \\\hline
{\color{blue}\bf $s\to u\ell^-\bar{\nu}_\ell:$}&\\
$\sqrt{2}H(\Xi^-\rightarrow\Sigma^0\ell^-\bar{\nu}_\ell)$&$-(a'_{31}+a'_{32}+a'_{33}+a'_{35})$&$-A_{31}$ \\
$\sqrt{6}H(\Xi^-\rightarrow\Lambda^0\ell^-\bar{\nu}_\ell)$&$a'_{31}+a'_{32}+a'_{33}+2a'_{34}-a'_{35}$&$A_{31}+2A_{32}$ \\
$H(\Xi^0\rightarrow\Sigma^+\ell^-\bar{\nu}_\ell)$&$a'_{31}+a'_{32}+a'_{33}+a'_{35}$&$A_{31}$  \\
$\sqrt{6}H(~~\Lambda^0\rightarrow p\ell^-\bar{\nu}_\ell~)$&$-(2a'_{31}+2a'_{32}+2a'_{33}+a'_{34}+a'_{35})$&$-(2A_{31}+A_{32})$  \\
$\sqrt{2}H(~~\Sigma^0\rightarrow p\ell^-\bar{\nu}_\ell~)$&$a'_{34}-a'_{35}$ &$A_{32}$\\
$H(~\Sigma^-\rightarrow n\ell^-\bar{\nu}_\ell~)$&$-(a'_{34}-a'_{35})$&$-A_{32}$ \\\hline
{\color{blue}\bf $d\to ue^-\bar{\nu}_e:$}&\\
$\sqrt{2}H(\Sigma^-\rightarrow \Sigma^0e^-\bar{\nu}_e)$&$-(a'_{21}+a'_{22}+a'_{23}+a'_{24})$&$-(2A_{21}-\bar{A}_{22})$ \\
$\sqrt{6}H(\Sigma^-\rightarrow \Lambda^0e^-\bar{\nu}_e)$&$a'_{21}+a'_{22}+a'_{23}-a'_{24}+2a'_{22}$&$\bar{A}_{22}$ \\
$\sqrt{2}H(\Sigma^0\rightarrow \Sigma^+e^-\bar{\nu}_e)$&$a'_{21}+a'_{22}+a'_{23}+a'_{24}$&$2A_{21}-\bar{A}_{22}$ \\
$\sqrt{6}H(\Sigma^+\rightarrow \Lambda^0e^+\nu_e)$&$a'_{21}+a'_{22}+a'_{23}-a'_{24}+2a'_{22}$&$\bar{A}_{22}$ \\
$H(\Xi^-\rightarrow \Xi^0e^-\bar{\nu}_e)$&$-a_{22}$&$\bar{A}_{22}-A_{21}$ \\
$H(~~n\rightarrow pe^-\bar{\nu}_e~~)$&$a'_{21}+a'_{22}+a'_{23}+a'_{22}$&$A_{21}$ \\\hline
\end{tabular}\label{Tab:ASLT8}}
\end{table}
The helicity amplitudes can be simplified by the redefinitions
\begin{eqnarray}
A_{n1}&=&a_{n1}+a_{n2}+a_{n3}+a_{n5},\nonumber\\
A_{n2}&=&a_{n4}-a_{n5}.
\end{eqnarray}
For convenience, we set $\bar{A}_{22}=A_{21}-A_{22}$ to replace $A_{22}$ for $d\to u\ell^-\bar{\nu}_\ell$ transition.  The reparameterization results are given in the last column of Table \ref{Tab:ASLT8}, in which we can easily see the helicity amplitude relations between different decay modes.

The differential branching ratios of $T_{8A} \to T_{8B} \ell^- \bar{\nu}_\ell$ decays can be written as
\begin{equation}
\frac{d\mathcal{B}(T_{8A} \to T_{8B} \ell^- \bar{\nu}_\ell)}{dq^2}=\frac{G_F^2|V_{uq_n}|^2\tau_{A}|\vec{p}_{B}|q^2}{192\pi^3m^2_{A}}\left(1-\frac{m_\ell^2}{q^2}\right)^2\left[B_{1}+\frac{m_\ell^2}{2q^2}B_{2}\right],\label{Eq:dBdsT8}
\end{equation}
with
\begin{eqnarray}
B_{1} &=& \Big|H_{\frac{1}{2} 0}\Big|^2+\Big|H_{-\frac{1}{2} 0}\Big|^2+\Big|H^2_{\frac{1}{2} 1}\Big|^2+\Big|H_{-\frac{1}{2} -1}\Big|^2,\nonumber\\
B_{2} &=& \Big|H_{\frac{1}{2} 0}\Big|^2+\Big|H_{-\frac{1}{2} 0}\Big|^2+\Big|H_{\frac{1}{2} 1}\Big|^2+\Big|H_{-\frac{1}{2} -1}\Big|^2+3\Big(\Big|H_{\frac{1}{2} t}\Big|^2+\Big|H_{-\frac{1}{2} t}\Big|^2\Big).\label{Eq:Bi}
\end{eqnarray}
The differential longitudinal branching ratios  $d\mathcal{B}^L(T_{8A}\to T_{8B} \ell^- \bar{\nu}_\ell)/dq^2$ can be obtained from $d\mathcal{B}(T_{8A} \to T_{8B} \ell^- \bar{\nu}_\ell)/dq^2$ by setting $\big|H^2_{\frac{1}{2} 1}\big|^2=\big|H_{-\frac{1}{2} -1}\big|^2=0$ in Eqs. (\ref{Eq:dBdsT8}-\ref{Eq:Bi}).

The theoretical input parameters and the experimental data within the $1\sigma$ error from  Particle Data Group  \cite{PDG2018} will be used in our numerical results. Two cases will be considered in our analysis:

\begin{itemize}
\item[\bf$S_1$:] Neglecting $B_2$ term in Eq. (\ref{Eq:dBdsT8}) as in Ref. \cite{Geng:2019bfz}   and treating $SU(3)$ flavor parameters $(a_{ni})^{V(A)}_{\lambda_B\lambda_{W}}(q^2)$ as constants without the $q^2$ dependence, $i.e.$ $B_1$ in Eq. (\ref{Eq:dBdsT8}) is constant. Then there are three parameters
\begin{eqnarray}
&&A_{31},A_{32}e^{i\delta_{A_{32}}}~~~~\mbox{for  $s\to u\ell^-\bar{\nu}_\ell$ transition}, \nonumber\\
&&A_{21},\bar{A}_{22}e^{i\delta_{\bar{A}_{22}}}~~~~ \mbox{for  $d\to ue^-\bar{\nu}_e$ transition}.
\end{eqnarray}
Noted that,  $A_{31}$, $A_{32}$, $A_{21}$ and $\bar{A}_{22}$ could be complex. In this work,  we set $A_{31}$($A_{21}$) is real and add relative phase $\delta_{A_{32}}$($\delta_{\bar{A}_{22}}$) associated with $A_{32}$($\bar{A}_{22}$).

For $s\to u\mu^-\bar{\nu}_\mu$ and $s\to ue^-\bar{\nu}_e$   transitions, firstly, we use the experimental measurements of $\mathcal{B}(\Xi^0\rightarrow\Sigma^+e^-\bar{\nu}_e)$ and $\mathcal{B}(\Sigma^-\rightarrow ne^-\bar{\nu}_e)$ to obtain $|A_{31}|$ and $|A_{32}|$ , secondly, we use the data of $\mathcal{B}(\Lambda^0\rightarrow pe^-\bar{\nu}_e)$ to constrain
$\delta_{A_{32}}$, which varies in the region $[-180^\circ,180^\circ]$, and then we give the predictions of relevant branching ratios.
For $d\to u e^-\bar{\nu}_e$  transition,  we use the experimental measurements of $\mathcal{B}(n\rightarrow pe^-\bar{\nu}_e)$ and $\mathcal{B}(\Sigma^-\rightarrow \Lambda^0e^-\bar{\nu}_e)$ to obtain $|A_{21}|$ and $|\bar{A}_{22}|$, and then let the predictions  satisfy other two experimental measurements.

\item[\bf $S_2$:] In order to obtain more precise predictions, we use the helicity
amplitudes in Eq. (\ref{Eq:HVA}). The form factors for the hyperon semileptonic decays are calculated in various approaches,  for examples, quark and soliton models, $1/N_c$ expansion of QCD, lattice QCD and  chiral perturbation
theory etc  \cite{Sasaki:2012ne,Sasaki:2008ha,Villadoro:2006nj,Lacour:2007wm,Faessler:2008ix,Guadagnoli:2006gj,Carrillo-Serrano:2014zta,Cabibbo:2003cu,Ledwig:2008ku,Faessler:2007pp,Yang:2015era}.
In this case,  we choose the dipole behavior for the form factors as  \cite{Gaillard:1984ny,Cabibbo:2003cu}
\begin{eqnarray}
F_i(q^2)={F_i(0)\over (1-q^2/M^2)^2},
\end{eqnarray}
where $M=0.97 ~(1.25)$  GeV for the vector (axial vector) form factors $f_i$ ($g_i$) in $s\to u\ell^-\bar{\nu}_\ell$ decays,   and $M=0.84\pm0.04~(1.08\pm0.08)$ GeV for $f_i$ ($g_i$) in $d\to ue^-\bar{\nu}_e$ decays.
For the form factor ratios $g_1(0)/f_1(0)$ and $f_2(0)/f_1(0)$,  they are preferentially taken from experimental measurements. If no relevant experimental measurements are available, they will be taken from Cabibbo theory \cite{Cabibbo:2003cu}.  The form factor ratios in Tab. \ref{Tab:R12figi} will be used in our results. As a result, the branching ratios only depend on the form factor $f_1(0)$ and  the CKM matrix elemant $V_{uq_n}$.
\begin{table}[htb]
\renewcommand\arraystretch{1.5}
\tabcolsep 0.35in
\centering
\caption{The form factor ratios $g_1(0)/f_1(0)$ and  $f_2(0)/f_1(0)$ from PDG2018 \cite{PDG2018} unless otherwise specified. $^a$denotes that the values are obtained from  the SU(3)-favour parametrization $F$ and $D$ given in Refs. \cite{Gaillard:1984ny,Cabibbo:2003cu} and the  measured  form  factor ratios in Ref. \cite{PDG2018}, and $^b$denotes that the values are taken from Cabibbo theory \cite{Cabibbo:2003cu}.}\vspace{0.1cm}
{\footnotesize
\begin{tabular}{lcccc}  \hline\hline
Decay modes  & $g_1(0)/f_1(0)$&$f_2(0)/f_1(0)$ \\\hline
$\Xi^-\rightarrow\Sigma^0\ell^-\bar{\nu}_\ell$&$1.22\pm0.05^a$&$2.609^b$\\
$\Xi^-\rightarrow\Lambda^0\ell^-\bar{\nu}_\ell$&$0.25\pm0.05$&$0.085^b$ \\
$\Xi^0\rightarrow\Sigma^+\ell^-\bar{\nu}_\ell$&$1.22\pm0.05$&$2.0\pm0.9$ \\
$\Lambda^0\rightarrow p\ell^-\bar{\nu}_\ell$&$0.718\pm0.015$&$1.066^b$ \\
$\Sigma^0\rightarrow p\ell^-\bar{\nu}_\ell$&$-0.340\pm0.017^a$&$-1.292^b$\\
$\Sigma^-\rightarrow n\ell^-\bar{\nu}_\ell$&$-0.340\pm0.017$&$-0.97\pm0.14$  \\\hline
$\Sigma^-\rightarrow \Sigma^0e^-\bar{\nu}_e$&$\frac{1}{2}\big[(1.2724\pm0.0023)+(-0.340\pm0.017)\big]^a$&$0.534^b$ \\
$\Sigma^-\rightarrow \Lambda^0e^-\bar{\nu}_e$&$(-0.01\pm0.10)^{-1}$&$1.490^b$ \\
$\Sigma^0\rightarrow \Sigma^+e^-\bar{\nu}_e$&$-\frac{1}{2}\big[(1.2724\pm0.0023)+(-0.340\pm0.017)\big]^a$&$0.531^b$  \\
$\Sigma^+\rightarrow \Lambda^0e^+\nu_e$&$(-0.01\pm0.10)^{-1a}$&$1.490^b$ \\
$\Xi^-\rightarrow \Xi^0e^-\bar{\nu}_e$&$-0.340\pm0.017^a$&$-1.432^b$ \\
$n\rightarrow pe^-\bar{\nu}_e$&$1.2724\pm0.0023$&$1.855^b$ \\\hline
\end{tabular}\label{Tab:R12figi}}
\end{table}
Then these three parameters become
\begin{eqnarray}
&&A'_{31},A'_{32}e^{i\delta_{A'_{32}}}~~~~\mbox{for  $s\to u\ell^-\bar{\nu}_\ell$ transition}, \nonumber\\
&&A'_{21},\bar{A}'_{22}e^{i\delta_{\bar{A}'_{22}}}~~~~ \mbox{for  $d\to ue^-\bar{\nu}_e$ transition},
\end{eqnarray}
where $A'_{ni}$  contains  $f_1(0)$  but without the $q^2$ dependence.
Finally,  all experimental data will be considered to constrain these parameters and predict the not-yet-measured branching ratios.
\end{itemize}

%
\begin{table}[t]
\renewcommand\arraystretch{1.5}
\tabcolsep 0.23in
\centering
\caption{The experimental data and the SM predictions with the $\pm1\sigma$ error bar of branching ratios of $T_{8A}\to T_{8B}\ell\nu_\ell$. $^\ddagger$denotes which experimental data give the finally effective constraints on the parameters, and $^\dagger$denotes the predictions depend  on the relative phase, which is not constrained well from present data.}\vspace{0.1cm}
{\footnotesize
\begin{tabular}{lcccc}  \hline\hline
~~~~~~~ Observables & Exp. Data \cite{PDG2018}& $\mathcal{B}r-S_1$ & $\mathcal{B}r-S_2$ & $\mathcal{B}r^L-S_2$ \\\hline
$\mathcal{B}(\Xi^-\rightarrow\Sigma^0e^-\bar{\nu}_e)(\times10^{-5})$&$8.7\pm1.7$&$8.12\pm0.60$&$8.27\pm0.58$ &$5.23\pm0.35$ \\
$\mathcal{B}(\Xi^-\rightarrow\Lambda^0e^-\bar{\nu}_e)(\times10^{-4})$&$5.63\pm0.31$&$1.21\pm0.71$&$5.47\pm0.15^\ddagger$&$4.94\pm0.14$  \\
$\mathcal{B}(\Xi^0\rightarrow\Sigma^+e^-\bar{\nu}_e)(\times10^{-4})$&$2.52\pm0.08$&$2.52\pm0.08^\ddagger$&$2.52\pm0.08^\ddagger$ &$1.60\pm0.06$ \\
$\mathcal{B}(\Lambda^0\rightarrow pe^-\bar{\nu}_e)(\times10^{-4})$&$8.32\pm0.14$&$8.32\pm0.14^\ddagger$&$8.32\pm0.14^\ddagger$ &$6.05\pm0.13$ \\
$\mathcal{B}(\Sigma^0\rightarrow pe^-\bar{\nu}_e)(\times10^{-13})$&$\cdots$&$2.41\pm0.32$&$2.46\pm0.32$ &$2.01\pm0.26$ \\
$\mathcal{B}(\Sigma^-\rightarrow ne^-\bar{\nu}_e)(\times10^{-3})$&$1.017\pm0.034$&$1.017\pm0.034^\ddagger$&$1.013\pm0.030^\ddagger$&$0.851\pm0.034$  \\\hline
$\mathcal{B}(\Xi^-\rightarrow\Sigma^0\mu^-\bar{\nu}_\mu)(\times10^{-6})$&$\leq800$&$1.08\pm0.09$&$1.13\pm0.08$&$0.57\pm0.04$  \\
$\mathcal{B}(\Xi^-\rightarrow\Lambda^0\mu^-\bar{\nu}_\mu)(\times10^{-4})$&$3.5^{+3.5}_{-2.2}$&$0.33\pm0.19$&$1.58\pm0.04$&$1.41\pm0.04$  \\
$\mathcal{B}(\Xi^0\rightarrow\Sigma^+\mu^-\bar{\nu}_\mu)(\times10^{-6})$&$2.33\pm0.35$&$2.14\pm0.14$&$2.18\pm0.1$&$1.09\pm0.08$  \\
$\mathcal{B}(\Lambda^0\rightarrow p\mu^-\bar{\nu}_\mu)(\times10^{-4})$&$1.57\pm0.35$&$1.35\pm0.02$&$1.40\pm0.02$&$0.94\pm0.02$  \\
$\mathcal{B}(\Sigma^0\rightarrow p\mu^-\bar{\nu}_\mu)(\times10^{-13})$&$\cdots$&$1.05\pm0.14$&$1.13\pm0.15$&$0.92\pm0.12$  \\
$\mathcal{B}(\Sigma^-\rightarrow n\mu^-\bar{\nu}_\mu)(\times10^{-4})$&$4.5\pm0.4$&$4.53\pm0.15$&$4.76\pm0.14^\ddagger$ &$3.99\pm0.17$ \\\hline
$\mathcal{B}(\Sigma^-\rightarrow \Sigma^0e^-\bar{\nu}_e)(\times10^{-10})$&$\cdots$&$4.36\pm4.01^\dag$&$1.35\pm0.28^\dag$&$1.11\pm0.23^\dag$  \\
$\mathcal{B}(\Sigma^-\rightarrow \Lambda^0e^-\bar{\nu}_e)(\times10^{-5})$&$5.73\pm0.27$&$5.73\pm0.27^\ddagger$&$5.73\pm0.27^\ddagger$&$3.18\pm0.15$  \\
$\mathcal{B}(\Sigma^0\rightarrow \Sigma^+e^-\bar{\nu}_e)(\times10^{-20})$&$\cdots$&$3.41\pm3.20^\dag$&$0.97\pm0.35^\dag$&$0.80\pm0.28^\dag$  \\
$\mathcal{B}(\Sigma^+\rightarrow \Lambda^0e^+\nu_e)(\times10^{-5})$&$2.0\pm0.5$&$1.88\pm0.11$&$1.86\pm0.11$ &$1.04\pm0.06$ \\
$\mathcal{B}(\Xi^-\rightarrow \Xi^0e^-\bar{\nu}_e)(\times10^{-9})$&$\leq2.3\times10^{6}$&$2.57\pm2.53^\dag$&$0.42\pm0.24^\dag$&$0.37\pm0.21^\dag$ \\
$\mathcal{B}(n\rightarrow pe^-\bar{\nu}_e)$&$100\%$&$100\%^\ddagger$&$100\%^\ddagger$&$(58.38\pm0.03)\%$ \\\hline
\end{tabular}\label{Tab:ASLNT8}}
\end{table}
Firstly,  we give a comment on the results of the twelve  $s\to u \ell^-\bar{\nu}_\ell$ decay modes.
In $S_1$ case, we get $A_{31}=5.87\pm0.21$, $A_{32}=2.57\pm0.06$, $|\delta_{A_{32}}|\leq155.90^\circ$ and the predictions are listed in the second column of Tab. \ref{Tab:ASLNT8}. One can see that when the branching ratio predictions satisfy the data of  $\mathcal{B}(\Xi^0\rightarrow\Sigma^+e^-\bar{\nu}_e)$, $\mathcal{B}(\Sigma^-\rightarrow ne^-\bar{\nu}_e)$ and $\mathcal{B}(\Lambda^0\rightarrow pe^-\bar{\nu}_e)$, the predictions of $\mathcal{B}(\Xi^-\rightarrow\Lambda^0e^-\bar{\nu}_e)$ and $\mathcal{B}(\Xi^-\rightarrow\Lambda^0\mu^-\bar{\nu}_\mu)$ obviously deviate from their experimental data.
In $S_2$ case, we consider  $q^2$-dependence of the form factors and  all relevant experimental constraints. We get $A'_{31}=1.04\pm0.04$, $A'_{32}=0.98\pm0.03$, $|\delta_{A'_{32}}|\leq28^\circ$, and the branching ratio predictions  are given  in the third column of Tab. \ref{Tab:ASLNT8}.  We can see that the experimental data of $\mathcal{B}(\Xi^-\rightarrow\Lambda^0e^-\bar{\nu}_e,~\Xi^0\rightarrow\Sigma^+e^-\bar{\nu}_e,~\Lambda^0\rightarrow pe^-\bar{\nu}_e,~\Sigma^-\rightarrow ne^-\bar{\nu}_e,\Sigma^-\rightarrow n\mu^-\bar{\nu}_\mu)$   give the finally effective constraints on the relevant parameters, and  the SU(3) IRA predictions in $S_2$ case are quite consistent with the present data within $1\sigma$ error.
We predict that $\mathcal{B}(\Xi^{-}\rightarrow \Sigma^0\mu^-\bar{\nu}_\mu)$ is at $10^{-6}$ order of magnitude, which is promising to be observed by the BESIII and LHCb experiments.

Then  we comment the results of the six  $d\to u e^-\bar{\nu}_e$ decay modes. Three branching  ratios $\mathcal{B}(\Sigma^-\rightarrow \Lambda^0e^-\bar{\nu}_e)$,  $\mathcal{B}(\Sigma^+\rightarrow \Lambda^0e^+\nu_e)$ and $\mathcal{B}(n\rightarrow pe^-\bar{\nu}_e)$ are precisely measured, which can be used to constrain on $A^{(')}_{21}$ and $\bar{A}^{(')}_{22}$ but not on the relative phase $\delta_{\bar{A}^{(')}_{22}}$, so we have quite large errors in the predictions of  $\mathcal{B}(\Sigma^-\rightarrow \Sigma^0e^-\bar{\nu}_e,\Sigma^0\rightarrow \Sigma^+e^-\bar{\nu}_e,\Xi^-\rightarrow \Xi^0e^-\bar{\nu}_e)$.
 We obtain $A_{21}=4.61\pm0.01$ and $\bar{A}_{22}=5.85\pm0.16$  in $S_1$ case as well as  $A'_{21}=4.50\pm0.02$ and $\bar{A}'_{22}=0.36\pm0.36$  in $S_2$ case.  The predictions for  $\mathcal{B}(\Sigma^-\rightarrow \Sigma^0e^-\bar{\nu}_e,\Sigma^0\rightarrow \Sigma^+e^-\bar{\nu}_e,\Xi^-\rightarrow \Xi^0e^-\bar{\nu}_e)$  in $S_2$ case are obviously different from that in $S_1$ case.
We  predict that  $\mathcal{B}(\Sigma^-\rightarrow \Sigma^0e^-\bar{\nu}_e,\Xi^-\rightarrow \Xi^0e^-\bar{\nu}_e)$  are at the order of $10^{-10}$ in $S_2$ case, which should be tested by the future experiments.

The  longitudinal branching ratios of  $T_{8A} \to T_{8B} \ell^- \bar{\nu}_\ell$  decays are also predicted in $S_2$ case, which are listed in the last column of Tab. \ref{Tab:ASLNT8}.  Noted that the life time of $\Sigma^0$ is very small, so the relevant decay branching ratios are also very small, and the same things happen in latter $\Xi^{*0,-}$, $\Sigma^{*0,-}$ and $\Delta^{0,-}$ semileptonic decays.

\subsection{$T_{10}\to T_{8}\ell^-\bar{\nu}_\ell$ semileptonic decays}
The feynman diagram for $T_{10}\to T_{8}\ell^-\bar{\nu}_\ell$ decays is also shown in Fig. \ref{fig:SL}.
Similar to $T_{8A}\to T_{8B}\ell^-\bar{\nu}_\ell$ semileptonic decays, the SU(3) IRA helicity amplitudes $H^{V(A)}_{\lambda_B \lambda_{W}}$ for  $T_{10}\to T_{8}\ell^-\bar{\nu}_\ell$ decays can be parameterized as
\begin{eqnarray}
H^{V(A),IRA}_{\lambda_B \lambda_{W}}&=&~~b_{n1}H(3)^k_n(T_{10})^{nij}(T_8)_{[ik]j},
\end{eqnarray}
with $b_{n1}\equiv(b_{n1})^{V(A)}_{\lambda_B\lambda_{W}}(q^2)$.    The  helicity amplitudes $H^{V(A)}_{\lambda_B \lambda_{W}}$ for different $T_{10}\to T_8\ell^-\bar{\nu}_\ell$ decays are given in Tab. \ref{Tab:T102T8lvH}.
\begin{table}[b]
\renewcommand\arraystretch{1.5}
\tabcolsep 0.4in
\centering
\caption{The helicity amplitudes $H^{V(A)}_{\lambda_B \lambda_{W}}$ of $T_{10}\to T_8\ell^-\bar{\nu}_\ell$ decays.}\vspace{0.1cm}
{\footnotesize
\begin{tabular}{rcc}  \hline\hline
$H^{V(A)}_{\lambda_B\lambda_W}$~~~~~~~~~~ &  SU(3) IRA amplitudes  \\\hline
{\color{blue}\bf $s\to u\ell^-\bar{\nu}_\ell:$}&\\
$H(\Omega^-\rightarrow\Xi^0\ell^-\bar{\nu}_\ell)$&$b_{31}$ \\
$3\sqrt{2}H(\Xi^{*-}\rightarrow\Lambda^0\ell^-\bar{\nu}_\ell)$&$3b_{31}$ \\
$\sqrt{6}H(\Xi^{*-}\rightarrow\Sigma^0\ell^-\bar{\nu}_\ell)$&$b_{31}$ \\
$\sqrt{3}H(\Xi^{*0}\rightarrow \Sigma^+\ell^-\bar{\nu}_\ell)$&$b_{31}$ \\
$\sqrt{3}H(~\Sigma^{*-}\rightarrow n\ell^-\bar{\nu}_\ell)$&$-b_{31}$ \\
$\sqrt{6}H(~\Sigma^{*0}\rightarrow p\ell^-\bar{\nu}_\ell~)$&$-b_{31}$ \\\hline
{\color{blue}\bf $d\to ue^-\bar{\nu}_e:$}&\\
$\sqrt{3}H(\Xi^{*-}\rightarrow \Xi^0e^-\bar{\nu}_e)$&$b_{21}$ \\
$2\sqrt{3}H(\Sigma^{*-}\rightarrow \Lambda^0e^-\bar{\nu}_e)$&$3b_{21}$  \\
$\sqrt{6}H(\Sigma^{*-}\rightarrow \Sigma^0e^-\bar{\nu}_e)$&$b_{21}$  \\
$\sqrt{6}H(\Sigma^{*0}\rightarrow \Sigma^+e^+\nu_e)$&$b_{21}$  \\
$H(~\Delta^-\rightarrow ne^-\bar{\nu}_e~)$&$-b_{21}$  \\
$\sqrt{3}H(~\Delta^0\rightarrow pe^-\bar{\nu}_e~~)$&$-b_{21}$  \\\hline
\end{tabular}\label{Tab:T102T8lvH}}
\end{table}
\begin{table}[t]
\renewcommand\arraystretch{1.5}
\tabcolsep 0.3in
\centering
\caption{The experimental data and the SU(3) IRA predictions with the $\pm1\sigma$ error bar of  $\mathcal{B}(T_{10}\to T_8\ell^-\bar{\nu}_\ell)$.}\vspace{0.1cm}
{\footnotesize
\begin{tabular}{lccc}  \hline\hline
 ~~~~~~~~Observables & Exp. Data \cite{PDG2018}& $S_1$   \\\hline
$\mathcal{B}(\Omega^-\rightarrow\Xi^0e^-\bar{\nu}_e)(\times10^{-3})$&$5.6\pm2.8$&$5.6\pm2.8^\ddag$ \\
$\mathcal{B}(\Xi^{*-}\rightarrow\Lambda^0e^-\bar{\nu}_e)(\times10^{-15})$&$\cdots$&$6.6\pm4.1$ \\
$\mathcal{B}(\Xi^{*-}\rightarrow\Sigma^0e^-\bar{\nu}_e)(\times10^{-15})$&$\cdots$&$2.2\pm1.4$ \\
$\mathcal{B}(\Xi^{*0}\rightarrow \Sigma^+e^-\bar{\nu}_e)(\times10^{-15})$&$\cdots$&$1.6\pm0.9$\\
$\mathcal{B}(\Sigma^{*-}\rightarrow ne^-\bar{\nu}_e)(\times10^{-15})$&$\cdots$&$1.6\pm0.9$ \\
$\mathcal{B}(\Sigma^{*0}\rightarrow pe^-\bar{\nu}_e)(\times10^{-16})$&$\cdots$&$9.3\pm5.5$\\\hline
$\mathcal{B}(\Omega^-\rightarrow\Xi^0\mu^-\bar{\nu}_\mu)(\times10^{-3})$&$\cdots$&$3.7\pm1.8$\\
$\mathcal{B}(\Xi^{*-}\rightarrow\Lambda^0\mu^-\bar{\nu}_\mu)(\times10^{-15})$&$\cdots$&$4.9\pm3.0$ \\
$\mathcal{B}(\Xi^{*-}\rightarrow\Sigma^0\mu^-\bar{\nu}_\mu)(\times10^{-15})$&$\cdots$&$1.6\pm1.0$\\
$\mathcal{B}(\Xi^{*0}\rightarrow \Sigma^+\mu^-\bar{\nu}_\mu)(\times10^{-15})$&$\cdots$&$1.0\pm0.5$ \\
$\mathcal{B}(\Sigma^{*-}\rightarrow n\mu^-\bar{\nu}_\mu)(\times10^{-15})$&$\cdots$&$1.2\pm0.7$\\
$\mathcal{B}(\Sigma^{*0}\rightarrow p\mu^-\bar{\nu}_\mu)(\times10^{-16})$&$\cdots$&$7.1\pm4.2$ \\\hline
$\mathcal{B}(\Xi^{*-}\rightarrow \Xi^0e^-\bar{\nu}_e)(\times10^{-15})$&$\cdots$&$3.6\pm2.2$ \\
$\mathcal{B}(\Sigma^{*-}\rightarrow \Lambda^0e^-\bar{\nu}_e)(\times10^{-15})$&$\cdots$&$6.2\pm3.3$ \\
$\mathcal{B}(\Sigma^{*-}\rightarrow \Sigma^0e^-\bar{\nu}_e)(\times10^{-16})$&$\cdots$&$2.7\pm1.4$ \\
$\mathcal{B}(\Sigma^{*0}\rightarrow \Sigma^+e^-\bar{\nu}_e)(\times10^{-16})$&$\cdots$&$3.1\pm1.8$\\
$\mathcal{B}(\Delta^-\rightarrow ne^-\bar{\nu}_e)(\times10^{-15})$&$\cdots$&$4.9\pm2.6$ \\
$\mathcal{B}(\Delta^0\rightarrow pe^-\bar{\nu}_e)(\times10^{-15})$&$\cdots$&$1.7\pm0.9$ \\\hline
\end{tabular}\label{Tab:T102T8lvNR}}
\end{table}
And the differential branching ratios of $T_{10A} \to T_{8B} \ell^- \bar{\nu}_\ell$ decays can be written as
\begin{equation}
\frac{d\mathcal{B}(T_{10A} \to T_{8B} \ell^- \bar{\nu}_\ell)}{dq^2}=\frac{G_F^2|V_{uq_n}|^2\tau_{A}|\vec{p}_{B}|q^2}{384\pi^3m^2_{A}}\left(1-\frac{m_\ell^2}{q^2}\right)^2\left[B'_1+\frac{m_\ell^2}{2q^2}B'_2\right],\label{Eq:dBds}
\end{equation}
with
\begin{eqnarray}
B'_1 &=& \Big|H_{\frac{1}{2} 0}\Big|^2+\Big|H_{-\frac{1}{2} 0}\Big|^2+\Big|H_{\frac{1}{2} 1}\Big|^2+\Big|H_{-\frac{1}{2} 1}\Big|^2+\Big|H_{-\frac{1}{2} -1}\Big|^2+\Big|H_{\frac{1}{2} -1}\Big|^2,\nonumber\\
B'_2 &=& \Big|H_{\frac{1}{2} 0}\Big|^2+\Big|H_{-\frac{1}{2} 0}\Big|^2+\Big|H_{\frac{1}{2} 1}\Big|^2+\Big|H_{-\frac{1}{2} 1}\Big|^2+\Big|H_{-\frac{1}{2} -1}\Big|^2+\Big|H_{\frac{1}{2} -1}\Big|^2+3\Big(\Big|H_{\frac{1}{2} t}\Big|^2+\Big|H_{-\frac{1}{2} t}\Big|^2\Big).
\end{eqnarray}
The $S_1$ case given in Sec. \ref{S88}  will be considered in $T_{10}\to T_{8}\ell^-\bar{\nu}_\ell$ semileptonic decays, where the $SU(3)_f$ parameters $(b_{n1})^{V(A)}_{\lambda_B\lambda_{W}}(q^2)$ are treated as constant without $q^2$-dependence.  The only  parameter is  $b_{31}$ for $s\to u \ell^-\bar{\nu}_\ell$ transition and $b_{21}$ for $d\to u e^-\bar{\nu}_e$ transition, respectively.

For $s\to u\ell^-\bar{\nu}_\ell$  transition, only $\mathcal{B}(\Omega^-\rightarrow\Xi^0e^-\bar{\nu}_e)$ has been measured. The experimental datum is listed in the second column of  Tab. \ref{Tab:T102T8lvNR}. We use  $\mathcal{B}(\Omega^-\rightarrow\Xi^0e^-\bar{\nu}_e)$ to constrain  $b_{31}$, and then give the predictions for other relevant decay branching ratios. The results are given in the third column of Tab. \ref{Tab:T102T8lvNR}. We obtain  $\mathcal{B}(\Omega^-\rightarrow\Xi^0\mu^-\bar{\nu}_\mu)=(3.7\pm1.8)\times10^{-3}$, which is very promising to be measured by the BESIII and LHCb.
For $d\to u e^-\bar{\nu}_e$  transition, no decay mode has been measured yet. We use $H(\Omega^-\rightarrow\Xi^0e^-\bar{\nu}_e)=H(\Delta^-\rightarrow ne^-\bar{\nu}_e)$ by the U-spin symmetry, $i.e.$ $b_{21}=-b_{31}$, to predict the branching ratios of $d\to u e^-\bar{\nu}_e$ transition, which are listed in the third column of Tab. \ref{Tab:T102T8lvNR}, too.
In Tab. \ref{Tab:T102T8lvNR}, all branching ratios except for $\mathcal{B}(\Omega^-\rightarrow\Xi^0e^-\bar{\nu}_e,\Xi^0\mu^-\bar{\nu}_\mu)$ are in the range of $10^{-16}-10^{-15}$, since the life times of  the  $\Xi^{*0,-}$, $\Sigma^{*0,-}$ and $\Delta^{0,-}$ baryons are very small.

\section{ Nonleptonic two-body decays of light baryons}
In this section, we discuss  the two-body nonleptonic decays of light baryons $T_{8,10}\to T_8 M_8$, where $M_8$ are
light pseudoscalar $P$ and vector $V$ meson  octets  under the $SU(3)$ flavor symmetry of $u,d,s$ quarks
\begin{eqnarray}
 P_8&=&\left(\begin{array}{ccc}
\frac{\eta_8}{\sqrt{6}}+\frac{\pi^0}{\sqrt{2}} & \pi^+ & K^+ \\
\pi^- &\frac{\eta_8}{\sqrt{6}}-\frac{\pi^0}{\sqrt{2}}  & K^0 \\
K^- & \bar{K}^0 &-\sqrt{\frac{2}{3}}\eta_8
\end{array}\right)\,,\\
V_8&=&\left(\begin{array}{ccc}
\frac{\omega_8}{\sqrt{6}}+\frac{\rho^0}{\sqrt{2}} & \rho^+ & K^{*+} \\
\rho^- &\frac{\omega_8}{\sqrt{6}}-\frac{\rho^0}{\sqrt{2}}  & K^{*0} \\
K^{*-} & \bar{K}^{*0} &-\sqrt{\frac{2}{3}}\omega_8
\end{array}\right)\,.
\end{eqnarray}

\subsection{ Weak decays of light baryons }

\begin{figure}[htbp]
\centering
\includegraphics[scale=0.9]{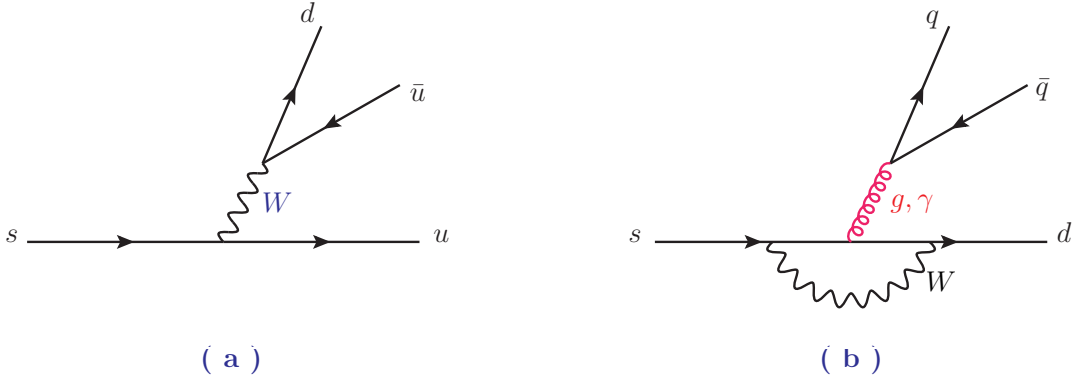}
\caption{Feynman diagrams for the $s$ quark decays in the SM. }
\label{fig:NLTreeP}
\end{figure}

In the SM, as shown in Fig. \ref{fig:NLTreeP}, there are two kinds of diagrams for the nonleptonic $s$ quark decays, the tree level diagram in Fig. \ref{fig:NLTreeP} (a) and the penguin diagram in Fig. \ref{fig:NLTreeP} (b).  The effective Hamiltonian for nonleptonic $s$ quark decays at scales $\mu<m_c$ can be written as \cite{Buchalla:1995vs}
\begin{eqnarray}
\mathcal{H}_{eff}=\frac{G_F}{\sqrt{2}}V_{ud}V_{us}^*\sum_{i=1}^{10}\Biggl[z_i(\mu)-\frac{V_{td}V_{ts}^*}{V_{ud}V_{us}^*}y_i(\mu)\Biggl]Q_i(\mu),\label{EQ:Heff}
\end{eqnarray}
where  $V_{uq}$ is the CKM matrix element,  $z_i(\mu)$ and $y_i(\mu)$ are Wilson coefficients. The four-quark operators $Q_i$ are
\begin{eqnarray}
&&Q_1=(\bar{d}_\alpha u_\beta)_{V-A}(\bar{u}_\beta s_\alpha)_{V-A},~~~~~~~~~~~~~~Q_2=(\bar{d}u)_{V-A}(\bar{u}s)_{V-A},\nonumber\\
&&Q_{3,5}=(\bar{d} s)_{V-A}\sum_{q=u,d,s} (\bar{q} q)_{V\mp A},~~~~~~~~~~Q_{4,6}=(\bar{d}_\beta s_\alpha)_{V-A}\sum_{q=u,d,s} (\bar{q}_\alpha q_\beta)_{V\mp A},\nonumber\\
&&Q_{7,9}=\frac{3}{2}(\bar{d} s)_{V-A}\sum_{q=u,d,s} e_q(\bar{q} q)_{V\pm A},~~~~~Q_{8,10}=\frac{3}{2}(\bar{d}_\beta s_\alpha)_{V-A}\sum_{q=u,d,s} e_q(\bar{q}_\alpha q_\beta)_{V\pm A},\label{EQ:Qi}
\end{eqnarray}
where $Q_{1,2}$ are current-current operators corresponding to Fig. \ref{fig:NLTreeP} (a), $Q_{3-6}$ ($Q_{7-10}$)  are QCD  (electroweak)  penguin operators corresponding to Fig. \ref{fig:NLTreeP} (b). In Eq. (\ref{EQ:Heff}), the magnetic penguin operators are ignored since their contributions  are small.
 $C_i(\mu)\equiv z_i(\mu)-\frac{V_{td}V_{ts}^*}{V_{ud}V_{us}^*}y_i(\mu)$  at $\mu=1$ GeV on $\Lambda^{(4)}_{\overline{MS}}$  in the NDR scheme are \cite{Buchalla:1995vs}
\begin{eqnarray}
&&C_1=-0.625,~~~C_2=1.361,~~~C_3=0.023,~~~C_4=-0.058,~~~C_5=0.009,~~~C_6=-0.059,\nonumber\\
&&C_7/\alpha_e=0.021,~~~C_8/\alpha_e=0.027,~~~ C_9/\alpha_e=0.036,~~~ C_{10}/\alpha_e=-0.015.\label{Eq:Ci}
\end{eqnarray}
Compared with tree-level contributions related to $C_{1,2}$, the penguin contributions are suppressed by smaller Wilson coefficients $C_{3,\cdots,10}$  and can be ignored in these decays.

The four-quark operators $Q_{i}$  can be rewritten as $(\bar{q}_iq^k)(\bar{q}_js)$ with $q_i=(u,d)$  as the doublet of 2 under the SU(2) symmetry by omitting the Lorentz-Dirac
structure. Since $(\bar{q}_iq^k)(\bar{q}_js)$   can be decomposed as the irreducible representations (IR) of  $(\bar{2}\otimes2\otimes\bar{2})s=(\bar{2}_p\oplus\bar{2}_t\oplus4)s$, one may obtain that
\begin{eqnarray}
\mathcal{O}(\bar{2}_p)_2&=&\big(\bar{u}u\big)\big(\bar{d}s\big)+\big(\bar{d}d\big)\big(\bar{d}s\big),\nonumber\\
\mathcal{O}(\bar{2}_t)_2&=&\big(\bar{d}u\big)\big(\bar{u}s\big)+\big(\bar{d}d\big)\big(\bar{d}s\big),\nonumber\\
\mathcal{O}(4)^1_{12}&=&\frac{1}{3}\big(\bar{u}u\big)\big(\bar{d}s\big)-\frac{1}{3}\big(\bar{d}d\big)\big(\bar{d}s\big)+\frac{1}{3}\big(\bar{d}u\big)\big(\bar{u}s\big),\label{Eq:Oi}
\end{eqnarray}
and we have the relation $ \mathcal{O}(4)^1_{21}=-\mathcal{O}(4)^2_{22}=\mathcal{O}(4)^1_{12}$ by the  traceless condition.  Then  $Q_{1,2}$, $Q_{3-6}$ and $Q_{7,10}$ can be transformed under SU(2) symmetry  as $\bar{2}_p\oplus\bar{2}_t\oplus4$, $\bar{2}_p/\bar{2}_t$ and $\bar{2}_p\oplus\bar{2}_t\oplus4$, respectively,
\begin{eqnarray}
Q_1&=&\mathcal{O}(4)^1_{12}+\frac{2}{3}\mathcal{O}(\bar{2}_p)_2-\frac{1}{3}\mathcal{O}(\bar{2}_t)_2,\nonumber\\
Q_2&=&\mathcal{O}(4)^1_{12}-\frac{1}{3}\mathcal{O}(\bar{2}_p)_2+\frac{2}{3}\mathcal{O}(\bar{2}_t)_2,\nonumber\\
Q_{3}&=&\mathcal{O}(\bar{2}_p)_2,~~~~~~~~~~~Q_{4}=\mathcal{O}(\bar{2}_t)_2,\nonumber\\
Q_{5}&=&\mathcal{O}'(\bar{2}_p)_2,~~~~~~~~~~Q_{6}=\mathcal{O}'(\bar{2}_t)_2,\nonumber\\
Q_{7}&=&\frac{3}{2}\mathcal{O}'(4)^1_{12}+\frac{1}{2}\mathcal{O}'(\bar{2}_p)_2-\frac{1}{2}\mathcal{O}'(\bar{2}_t)_2,\nonumber\\
Q_{8}&=&\frac{3}{2}\mathcal{O}'(4)^1_{12}-\frac{1}{2}\mathcal{O}'(\bar{2}_p)_2+\frac{1}{2}\mathcal{O}'(\bar{2}_t)_2,\nonumber\\
Q_{9}&=&\frac{3}{2}\mathcal{O}(4)^1_{12}+\frac{1}{2}\mathcal{O}(\bar{2}_p)_2-\frac{1}{2}\mathcal{O}(\bar{2}_t)_2,\nonumber\\
Q_{10}&=&\frac{3}{2}\mathcal{O}(4)^1_{12}-\frac{1}{2}\mathcal{O}(\bar{2}_p)_2+\frac{1}{2}\mathcal{O}(\bar{2}_t)_2,\label{Eq:RQiOi}
\end{eqnarray}
where $\mathcal{O}'(\bar{2}_p)_2$, $\mathcal{O}'(\bar{2}_t)_2$ and $\mathcal{O}'(4)^1_{12}$ are operators related to $Q_{5,6,7,8}$, which have the same SU(3) structure as $\mathcal{O}(\bar{2}_p)_2$, $\mathcal{O}(\bar{2}_t)_2$ and $\mathcal{O}(4)^1_{12}$ but different Lorentz-Dirac structures.

By using the bases of the SU(2) symmetry,
the effective Hamiltonian in Eq. (\ref{EQ:Heff}) can be transformed as
\begin{eqnarray}
\mathcal{H}_{eff}^{IR}&=&\frac{G_F}{\sqrt{2}}V_{ud}V_{us}^*\Bigg[
\bar{C}_4\mathcal{O}(4)^1_{12}+\bar{C}_{\bar{2}_p}\mathcal{O}(\bar{2}_p)_2+\bar{C}_{\bar{2}_t}\mathcal{O}(\bar{2}_t)_2+\bar{C}'_4\mathcal{O}'(4)^1_{12}
+\bar{C}'_{\bar{2}_p}\mathcal{O}'(\bar{2}_p)_2+\bar{C}'_{\bar{2}_t}\mathcal{O}'(\bar{2}_t)_2\Bigg], \label{Eq:HeffIR}
\end{eqnarray}
with
\begin{eqnarray}
\bar{C}_4&=&C_1+C_2+\frac{3}{2}\Big(C_9+C_{10}\Big),\nonumber\\
\bar{C}_{\bar{2}_p}&=&\frac{2}{3}C_1-\frac{1}{3}C_2+C_3+\frac{1}{2}\Big(C_9-C_{10}\Big),\nonumber\\
\bar{C}_{\bar{2}_t}&=&-\frac{1}{3}C_1+\frac{2}{3}C_2+C_4-\frac{1}{2}\Big(C_9-C_{10}\Big),\nonumber\\
\bar{C}'_{4}&=&\frac{3}{2}\Big(C_7+C_{8}\Big),\nonumber\\
\bar{C}'_{\bar{2}_p}&=&C_5+\frac{1}{2}\Big(C_7-C_{8}\Big),\nonumber\\
\bar{C}'_{\bar{2}_t}&=&C_6-\frac{1}{2}\Big(C_7-C_{8}\Big).
\end{eqnarray}

From Eq. (\ref{Eq:Ci}), one can see that the  contributions from  current-current operators related to $C_{1,2}$ are much larger than others related to $C_{3,\cdots,10}$. So we will only consider current-current operator contributions in the following analysis.
After neglecting $C_{3,\cdots,10}$, the effective Hamiltonian in Eq. (\ref{Eq:HeffIR}) can be rewritten as
\begin{eqnarray}
\mathcal{H}_{eff}^{IR}&=&\frac{G_F}{\sqrt{2}}V_{ud}V_{us}^*\Bigg\{
C_+\Bigg[2H(4)+\frac{1}{3}\Big(H(\bar{2}_t)+H(\bar{2}_p)\Big)\Bigg]+C_-\Big(H(\bar{2}_t)-H(\bar{2}_p)\Big)\Bigg\},\label{Eq:HeffIR2}
\end{eqnarray}
where $C_\pm\equiv (C_2\pm C_1)/2$, and $H^{ij}_k$ is  related to  $(\bar{q}_iq^k)(\bar{q}_js)$ operators.
From Eq. (\ref{Eq:Ci}), one gets $C^2_+/C^2_-\approx 13.7\%$, so $C_-$ term related to $H(\bar{2}_t)-H(\bar{2}_p)$ gives  the dominant contribution to the decay branching ratios.
 The non-zero entries of $H^{ij}_k$ corresponding to current-current operators in SU(2) flavor space are
\begin{eqnarray}
H(\bar{2}_p)^2=H(\bar{2}_t)^2=1, ~~~~H(4)^{12}_{1}=H(4)^{21}_{1}=\frac{1}{3}.\label{Eq:Hs1s}
\end{eqnarray}
Noted that $H(4)^{22}_{2}=-\frac{1}{3}$ only contributes to the penguin operators and we ignore it.

In Eq. (\ref{Eq:HeffIR2}), the  $\bar{2}$ irreducible representation is  linear combinations
of $\bar{2}_{p,t}$,  so we need only consider a single $\bar{2}$ when computing amplitudes from the
invariants and reduced matrix elements \cite{Grossman:2012ry}.

The amplitudes of the $T_{8,10}\to T_{8}M_8$ decays can be written via the effective Hamiltonian in Eq.  (\ref{EQ:Heff}) as
\begin{eqnarray}
A(T_{8,10}\to T_{8}M_8)=\langle T_{8}M_8|\mathcal{H}_{eff}|T_{8,10}\rangle. \label{Eq:A}
\end{eqnarray}
These amplitudes may be divided into the S wave   and  P wave  amplitudes, which have been analysed,  for instance, in  heavy baryon chiral perturbation
theory  \cite{Jenkins:1991bt,AbdElHady:1999mj,Flores-Mendieta:2019lao,Tandean:2002vy} and by using
a relativistic chiral unitary approach based on coupled channels \cite{Borasoy:2003rc}.
 Moreover, since  $\mathcal{H}_{eff}^{IR}$ is irreducible in the SU(2) symmetry, and the initial and final state baryons ($T_{8}$, $T_{10}$, $M_8$) are irreducible in the SU(3) symmetry, the amplitudes of $T_{8,10}\to T_{8}M_8$ can be further written as
\begin{eqnarray}
A(T_{8,10}\to T_{8}M_8)=\langle T_{8}M_8|\mathcal{H}_{eff}^{IR}|T_{8,10}\rangle=A(\mathcal{O}_{4})+A(\mathcal{O}_{\bar{2}}). \label{Eq:AIR}
\end{eqnarray}

\subsubsection{$T_{8}\to T_{8}M_8$ weak decays}

\begin{figure}[b]
\centering
\includegraphics[scale=1.05]{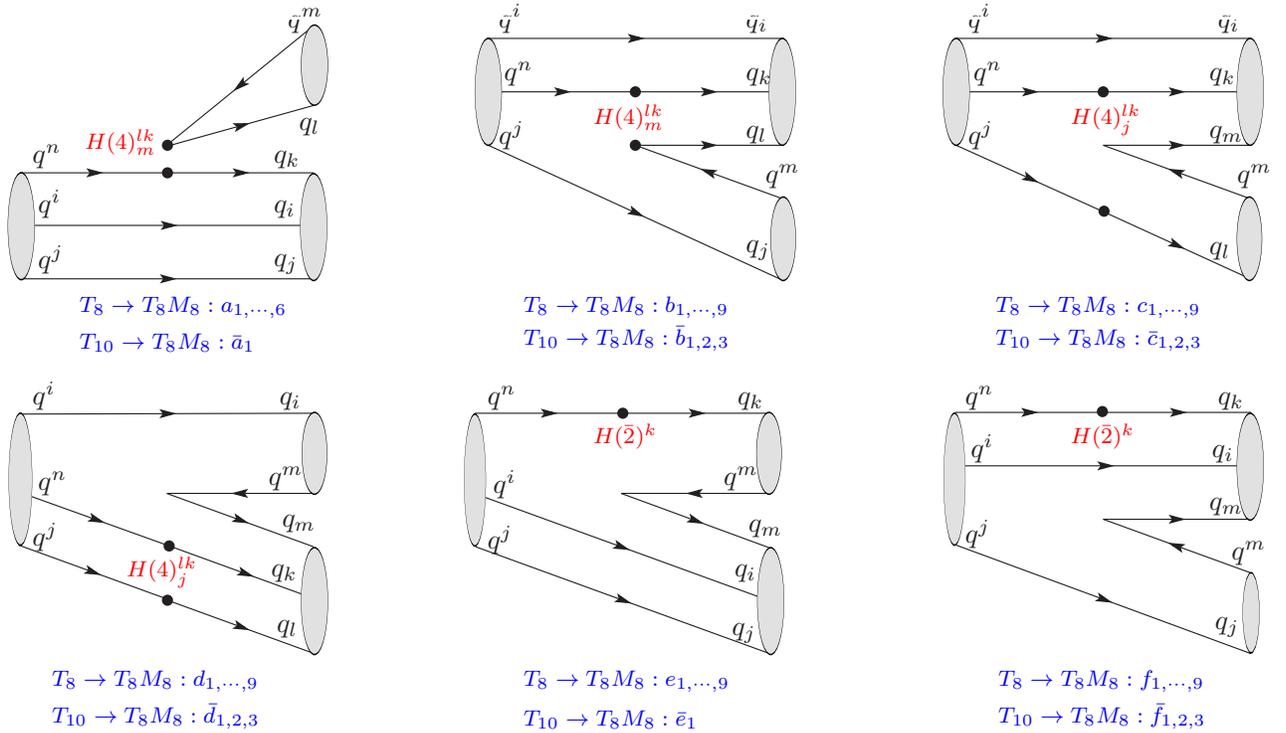}
\caption{Feynman diagrams of IRA  for $T_{8,10}\to T_{8}M_8$ nonleptonic  two-body  decays with $q^n=s$. }
\label{fig:NLSU3}
\end{figure}
Following Ref. \cite{Wang:2017azm}, the Feynman diagrams for $T_{8}\to T_{8}M_8$ nonleptonic $s$ quark decays are displayed in Fig. \ref{fig:NLSU3},
 and the SU(3) IRA amplitudes are
\newpage
{\small
\begin{eqnarray}
A(T_8\to T_8 M_8)^{IRA}&=&a_1H(4)^{lk}_m(T_8)^{[ij]n}(T_8)_{[ij]k}(M_8)^m_l+a_2H(4)^{lk}_m(T_8)^{[ij]n}(T_8)_{[ik]j}(M_8)^m_l+a_3H(4)^{lk}_m(T_8)^{[ij]n}(T_8)_{[jk]i}(M_8)^m_l\nonumber\\
&&+a_4H(4)^{lk}_m(T_8)^{[in]j}(T_8)_{[ij]k}(M_8)^m_l+a_5H(4)^{lk}_m(T_8)^{[in]j}(T_8)_{[ik]j}(M_8)^m_l
         +a_6H(4)^{lk}_m(T_8)^{[in]j}(T_8)_{[jk]i}(M_8)^m_l \nonumber\\
&&+a_7H(4)^{lk}_m(T_8)^{[jn]i}(T_8)_{[ij]k}(M_8)^m_l+a_8H(4)^{lk}_m(T_8)^{[jn]i}(T_8)_{[ik]j}(M_8)^m_l
         +a_9H(4)^{lk}_m(T_8)^{[jn]i}(T_8)_{[jk]i}(M_8)^m_l \nonumber\\
         &&+b_1H(4)^{lk}_m(T_8)^{[ij]n}(T_8)_{[il]k}(M_8)^m_j+b_2H(4)^{lk}_m(T_8)^{[ij]n}(T_8)_{[kl]i}(M_8)^m_j
          +b_3H(4)^{lk}_m(T_8)^{[ij]n}(T_8)_{[ki]l}(M_8)^m_j \nonumber\\
         && +b_4H(4)^{lk}_m(T_8)^{[in]j}(T_8)_{[il]k}(M_8)^m_j+b_5H(4)^{lk}_m(T_8)^{[in]j}(T_8)_{[kl]i}(M_8)^m_j
          +b_6H(4)^{lk}_m(T_8)^{[in]j}(T_8)_{[ki]l}(M_8)^m_j \nonumber\\
         && +b_7H(4)^{lk}_m(T_8)^{[jn]i}(T_8)_{[il]k}(M_8)^m_j+b_8H(4)^{lk}_m(T_8)^{[jn]i}(T_8)_{[kl]i}(M_8)^m_j
          +b_9H(4)^{lk}_m(T_8)^{[jn]i}(T_8)_{[ki]l}(M_8)^m_j \nonumber\\
         &&+c_1H(4)^{lk}_j(T_8)^{[ij]n}(T_8)_{[ki]m}(M_8)^m_l+c_2H(4)^{lk}_j(T_8)^{[ij]n}(T_8)_{[km]i}(M_8)^m_l
          +c_3H(4)^{lk}_j(T_8)^{[ij]n}(T_8)_{[im]k}(M_8)^m_l \nonumber\\
         && +c_4H(4)^{lk}_j(T_8)^{[in]j}(T_8)_{[ki]m}(M_8)^m_l+c_5H(4)^{lk}_j(T_8)^{[in]j}(T_8)_{[km]i}(M_8)^m_l
          +c_6H(4)^{lk}_j(T_8)^{[in]j}(T_8)_{[im]k}(M_8)^m_l \nonumber\\
         && +c_7H(4)^{lk}_j(T_8)^{[jn]i}(T_8)_{[ki]m}(M_8)^m_l+c_8H(4)^{lk}_j(T_8)^{[jn]i}(T_8)_{[km]i}(M_8)^m_l
          +c_9H(4)^{lk}_j(T_8)^{[jn]i}(T_8)_{[im]k}(M_8)^m_l \nonumber\\
         &&+d_1H(4)^{lk}_j(T_8)^{[ij]n}(T_8)_{[mk]l}(M_8)^m_i+d_2H(4)^{lk}_j(T_8)^{[ij]n}(T_8)_{[ml]k}(M_8)^m_i
          +d_3H(4)^{lk}_j(T_8)^{[ij]n}(T_8)_{[kl]m}(M_8)^m_i \nonumber\\
         && +d_4H(4)^{lk}_j(T_8)^{[in]j}(T_8)_{[mk]l}(M_8)^m_i+d_5H(4)^{lk}_j(T_8)^{[in]j}(T_8)_{[ml]k}(M_8)^m_i
          +d_6H(4)^{lk}_j(T_8)^{[in]j}(T_8)_{[kl]m}(M_8)^m_i \nonumber\\
         && +d_7H(4)^{lk}_j(T_8)^{[jn]i}(T_8)_{[mk]l}(M_8)^m_i+d_8H(4)^{lk}_j(T_8)^{[jn]i}(T_8)_{[ml]k}(M_8)^m_i
          +d_9H(4)^{lk}_j(T_8)^{[jn]i}(T_8)_{[kl]m}(M_8)^m_i \nonumber\\
         &&+e_1H(\bar{2})^k(T_8)^{[ij]n}(T_8)_{[ij]m}(M_8)^m_k+e_2H(\bar{2})^k(T_8)^{[ij]n}(T_8)_{[im]j}(M_8)^m_k
          +e_3H(\bar{2})^k(T_8)^{[ij]n}(T_8)_{[jm]i}(M_8)^m_k \nonumber\\
         && +e_4H(\bar{2})^k(T_8)^{[in]j}(T_8)_{[ij]m}(M_8)^m_k+e_5H(\bar{2})^k(T_8)^{[in]j}(T_8)_{[im]j}(M_8)^m_k
          +e_6H(\bar{2})^k(T_8)^{[in]j}(T_8)_{[jm]i}(M_8)^m_k \nonumber\\
        && +e_7H(\bar{2})^k(T_8)^{[jn]i}(T_8)_{[ij]m}(M_8)^m_k+e_8H(\bar{2})^k(T_8)^{[jn]i}(T_8)_{[im]j}(M_8)^m_k
          +e_9H(\bar{2})^k(T_8)^{[jn]i}(T_8)_{[jm]i}(M_8)^m_k \nonumber\\
          &&+f_1H(\bar{2})^k(T_8)^{[ij]n}(T_8)_{[ki]m}(M_8)^m_j+f_2H(\bar{2})^k(T_8)^{[ij]n}(T_8)_{[km]i}(M_8)^m_j
          +f_3H(\bar{2})^k(T_8)^{[ij]n}(T_8)_{[im]k}(M_8)^m_j \nonumber\\
         && +f_4H(\bar{2})^k(T_8)^{[in]j}(T_8)_{[ki]m}(M_8)^m_j+f_5H(\bar{2})^k(T_8)^{[in]j}(T_8)_{[km]i}(M_8)^m_j
          +f_6H(\bar{2})^k(T_8)^{[in]j}(T_8)_{[im]k}(M_8)^m_j \nonumber\\
         && +f_7H(\bar{2})^k(T_8)^{[jn]i}(T_8)_{[ki]m}(M_8)^m_j+f_8H(\bar{2})^k(T_8)^{[jn]i}(T_8)_{[km]i}(M_8)^m_j
          +f_9H(\bar{2})^k(T_8)^{[jn]i}(T_8)_{[im]k}(M_8)^m_j, \nonumber\\ \label{Eq:T82T8M8TRA}
         \end{eqnarray}}
where the coefficients $a_i,b_i,c_i,d_i,e_i,f_i$ are constants which contain the Wilson coefficients, CKM  matrix elements and information about QCD dynamics.
Using  $H(4)^{ab}_c$ is symmetric in upper indices, $b_i$ and $d_i$ terms  can be  simplified  by
\begin{eqnarray}
&&b_3=-b_1,~~~b_6=-b_4,~~~b_9=-b_7,~~~b_2=b_5=b_8=0,\nonumber\\
&&d_2=d_1,~~~d_5=d_4,~~~d_8=d_7,~~~d_3=d_6=d_9=0.
\end{eqnarray}
In addition,  using $i,j$ antisymmetric in $T_8^{[ij]n}$ and   $i,j$ indices  are arbitrary in $e_i$ terms, we have
\begin{eqnarray}
&&a_3=-a_2,~~~a_7=-a_4,~~~a_8=a_6,~~~a_9=a_5,~~~e_3=-e_2,~~~e_7=-e_4,~~~e_8=e_6,~~~e_9=e_5.
\end{eqnarray}
Finally, Eq.  (\ref{Eq:T82T8M8TRA}) can be simplified as
{\small
\begin{eqnarray}
A(T_8\to T_8 M_8)^{IRA,J}&=&a_1H(4)^{lk}_m(T_8)^{[ij]n}(T_8)_{[ij]k}(M_8)^m_l+a_2H(4)^{lk}_m(T_8)^{[ij]n}(T_8)_{[ik]j}(M_8)^m_l\nonumber\\
&&+a_4H(4)^{lk}_m(T_8)^{[in]j}(T_8)_{[ij]k}(M_8)^m_l+a_5H(4)^{lk}_m(T_8)^{[in]j}(T_8)_{[ik]j}(M_8)^m_l
         +a_6H(4)^{lk}_m(T_8)^{[in]j}(T_8)_{[jk]i}(M_8)^m_l \nonumber\\
         &&+b_1H(4)^{lk}_m(T_8)^{[ij]n}(T_8)_{[il]k}(M_8)^m_j+b_4H(4)^{lk}_m(T_8)^{[in]j}(T_8)_{[il]k}(M_8)^m_j +b_7H(4)^{lk}_m(T_8)^{[jn]i}(T_8)_{[il]k}(M_8)^m_j \nonumber\\
         &&+c_1H(4)^{lk}_j(T_8)^{[ij]n}(T_8)_{[ki]m}(M_8)^m_l+c_2H(4)^{lk}_j(T_8)^{[ij]n}(T_8)_{[km]i}(M_8)^m_l
          +c_3H(4)^{lk}_j(T_8)^{[ij]n}(T_8)_{[im]k}(M_8)^m_l \nonumber\\
        && +c_4H(4)^{lk}_j(T_8)^{[in]j}(T_8)_{[ki]m}(M_8)^m_l+c_5H(4)^{lk}_j(T_8)^{[in]j}(T_8)_{[km]i}(M_8)^m_l
          +c_6H(4)^{lk}_j(T_8)^{[in]j}(T_8)_{[im]k}(M_8)^m_l \nonumber\\
          && +c_7H(4)^{lk}_j(T_8)^{[jn]i}(T_8)_{[ki]m}(M_8)^m_l+c_8H(4)^{lk}_j(T_8)^{[jn]i}(T_8)_{[km]i}(M_8)^m_l
          +c_9H(4)^{lk}_j(T_8)^{[jn]i}(T_8)_{[im]k}(M_8)^m_l \nonumber\\
         &&+d_1H(4)^{lk}_j(T_8)^{[ij]n}(T_8)_{[mk]l}(M_8)^m_i+d_4H(4)^{lk}_j(T_8)^{[in]j}(T_8)_{[mk]l}(M_8)^m_i+d_7H(4)^{lk}_j(T_8)^{[jn]i}(T_8)_{[mk]l}(M_8)^m_i \nonumber\\
         &&+e_1H(\bar{2})^k(T_8)^{[ij]n}(T_8)_{[ij]m}(M_8)^m_k+e_2H(\bar{2})^k(T_8)^{[ij]n}(T_8)_{[im]j}(M_8)^m_k\nonumber\\
         && +e_4H(\bar{2})^k(T_8)^{[in]j}(T_8)_{[ij]m}(M_8)^m_k+e_5H(\bar{2})^k(T_8)^{[in]j}(T_8)_{[im]j}(M_8)^m_k
          +e_6H(\bar{2})^k(T_8)^{[in]j}(T_8)_{[jm]i}(M_8)^m_k \nonumber\\
          &&+f_1H(\bar{2})^k(T_8)^{[ij]n}(T_8)_{[ki]m}(M_8)^m_j+f_2H(\bar{2})^k(T_8)^{[ij]n}(T_8)_{[km]i}(M_8)^m_j
          +f_3H(\bar{2})^k(T_8)^{[ij]n}(T_8)_{[im]k}(M_8)^m_j \nonumber\\
         && +f_4H(\bar{2})^k(T_8)^{[in]j}(T_8)_{[ki]m}(M_8)^m_j+f_5H(\bar{2})^k(T_8)^{[in]j}(T_8)_{[km]i}(M_8)^m_j
          +f_6H(\bar{2})^k(T_8)^{[in]j}(T_8)_{[im]k}(M_8)^m_j\nonumber\\
           && +f_7H(\bar{2})^k(T_8)^{[jn]i}(T_8)_{[ki]m}(M_8)^m_j+f_8H(\bar{2})^k(T_8)^{[jn]i}(T_8)_{[km]i}(M_8)^m_j
          +f_9H(\bar{2})^k(T_8)^{[jn]i}(T_8)_{[im]k}(M_8)^m_j. \nonumber\\
        \label{Eq:T82T8M8TRA}
         \end{eqnarray}}

In Tab. \ref{Tab:T82T8M8IRA}, we list the IRA amplitudes of  $T_{8}\to T_{8}P_8$ weak decays, which include the $H(4)^{12}_1$, $H(4)^{22}_2$ and $H(\bar{2})^{2}$ terms. The corresponding $T_{8}\to T_{8}V_8$ weak decays have the same relations as $T_{8}\to T_{8}P_8$ weak decays.  If only considering the dominant contributions from $H(\bar{2})^2$  and redefining the parameters \begin{eqnarray}
A_1&=&2(e_5+e_6)+(f_4+f_5+f_7+f_8),\nonumber\\
A_2&=&2(e_5+e_6)+(f_5+f_6+f_8+f_9),\nonumber\\
A_3&=&2(e_5+e_6)-(f_4-f_5+f_7-f_8),\nonumber\\
A_4&=&4(e_1+e_2+e_4)+2(e_5-e_6)-(2f_1+2f_2+f_4+f_5-f_7-f_8),\nonumber\\
A_5&=&2(e_1+e_2+e_4+2e_5+e_6)-(f_1+2f_2+f_3)+(f_7+2f_8+f_9),
\end{eqnarray}
the IRA amplitudes can be greatly simplified as listed in the last  column of Tab. \ref{Tab:T82T8M8IRA}, in which we can easily see the relations of different decay amplitudes.

\begin{sidewaystable}
\renewcommand\arraystretch{3}
\tabcolsep 0.06in
\centering
\caption{The $SU(3)$ IRA amplitudes of $T_8\to T_8 P$ weak decays.   }\vspace{0.1cm}
{
\scriptsize
\begin{tabular}
{r|l|l|l|c}   \hline\hline
 Amplitudes~~~~~ &$\hspace{2.7cm}H(4)^{12}_{1}$ &$\hspace{2.7cm}H(4)^{22}_{2}$ &$\hspace{1.8cm}H(\bar{2})^{2}$ & Simplified Amplitudes   \\\hline
$\sqrt{2}A(\Sigma^+\rightarrow p\pi^0)$&\makecell[l]{ $-2(a_5+a_6)-2(b_4+b_7)$\\$+(c_4+2c_5+c_6)+(c_7+2c_8+c_9)-2(d_4+d_7)$}&$2(a_5+a_6)$&$2(e_5+e_6)+f_4+f_5+f_7+f_8$&$A_1$\\\hline
$A(\Sigma^+\rightarrow n\pi^+)$&$(c_4-c_6+c_7-c_9)+2(d_4+d_7)$&$-2(b_4+b_7)$&$f_4-f_6+f_7-f_9$&$A_1-A_2$\\\hline
$A(\Sigma^-\rightarrow n\pi^-)$&$-2(a_5+a_6)-2(b_4+b_7)$&$-(c_5+c_6+c_8+c_9)+2(d_4+d_7)$&$-2(e_5+e_6)-(f_5+f_6+f_8+f_9)$&$-A_2$\\\hline
$\sqrt{2}A(\Sigma^0\rightarrow p\pi^-)$&\makecell[l]{$2(a_5+a_6)-2(b_4+b_7)$\\$-(c_4-c_6+c_7-c_9)-2(d_4+d_7)$}&$c_4+c_5+c_7+c_8$&$2(e_5+e_6)+f_4+f_5+f_7+f_8$&$A_1$\\\hline
$2A(\Sigma^0\rightarrow n\pi^0)$&\makecell[l]{$-2(a_5+a_6)+2(b_4+b_7)$\\$+(c_4+2c_5+c_6)+(c_7+2c_8+c_9)-2(d_4+d_7)$}&\makecell[l]{$2(a_5+a_6)+2(b_4+b_7)$\\$-(c_4-c_6+c_7-c_9)-2(d_4+d_7)$}&$2(e_5+e_6)-(f_4-f_5+f_7-f_8)$&$A_3$\\\hline
$\sqrt{6}A(\Lambda^0\rightarrow p\pi^-)$&\makecell[l]{$-4(a_1+a_2+a_4)-2(a_5-a_6)-2(2b_1+b_4-b_7)$\\$+(2c_1-2c_3+c_4-c_6-c_7+c_9)+2(2d_1+d_4-d_7)$}&$2c_1+2c_2+c_4+c_5-c_7-c_8$&\makecell[l]{$-4(e_1+e_2+e_4)-2(e_5-e_6)$\\$+(2f_1+2f_2+f_4+f_5-f_7-f_8)$}&$-A_4$\\\hline
$2\sqrt{3}A(\Lambda^0\rightarrow n\pi^0)$&\makecell[l]{$-4(a_1+a_2+a_4)-2(a_5-a_6)-2(2b_1+b_4-b_7)$\\$-2(c_1+2c_2+c_3)-(c_4+2c_5+c_6)$\\$+(c_7+2c_8+c_9)+2(2d_1+d_4-d_7)$}&\makecell[l]{$4(a_1+a_2+a_4)+2(a_5-a_6)+2(2b_1+b_4-b_7)$\\$(-2c_1+2c_3-c_4+c_6+c_7-c_9)-2(2d_1+d_4-d_7)$}&\makecell[l]{$4(e_1+e_2+e_4)+2(e_5-e_6)$\\$-(2f_1+2f_2+f_4+f_5-f_7-f_8)$}&$A_4$\\\hline
$\sqrt{6}A(\Xi^-\rightarrow \Lambda^0\pi^-)$&$2(a_1+a_2+a_4+2a_5+a_6)$&$-(c_1+2c_2+c_3)+(c_7+2c_8+c_9)$&\makecell[l]{$2(e_1+e_2+e_4+2e_5+e_6)$\\$-(f_1+2f_2+f_3)+(f_7+2f_8+f_9)$}&$A_5$\\\hline
$2\sqrt{3}A(\Xi^0\rightarrow \Lambda^0\pi^0)$&\makecell[l]{$2(a_1+a_2+a_4+2a_5+a_6)$\\$2(c_1+2c_2+c_3)-2(c_7+2c_8+c_9)$}&$-2(a_1+a_2+a_4+2a_5+a_6)$&\makecell[l]{$-2(e_1+e_2+e_4+2e_5+e_6)$\\$+(f_1+2f_2+f_3)-(f_7+2f_8+f_9)$}&$-A_5$\\\hline
\end{tabular}\label{Tab:T82T8M8IRA}}
\end{sidewaystable}

The branching ratios of $T_8\to T_8 P_8$ can be written as
\begin{eqnarray}
\mathcal{B}(T_{8A}\to T_{8B} P_8)=\frac{\tau_A|p_{cm}|}{8\pi m_A^2}\big|A(T_{8A}\to T_{8B} P_8)\big|^2.
\end{eqnarray}
For more accurate results, we will consider the mass difference in the amplitudes \cite{He:2018joe}
\begin{eqnarray}
A(T_{8A}\to T_{8B} P_8)\propto\frac{m_A}{m_B}p_{cm}N_BN_A,\label{Eq:AMDT8}
\end{eqnarray}
with
\begin{eqnarray}
p_{cm}&=&\frac{1}{2m_A}\sqrt{\big(m_A^2-(m_B+m_P)^2\big)\big(m_A^2-(m_B-m_P)^2\big)},\nonumber\\
N_A&=&\sqrt{2m_A},\nonumber\\
N_B&=&\sqrt{\frac{(m_A+m_B)^2-m_P^2}{2m_A}}.
\end{eqnarray}

\begin{table}[t]
\renewcommand\arraystretch{1.5}
\tabcolsep 0.3in
\centering
\caption{The experimental measurements and the SM predictions with the $\pm1\sigma$ error bar of  branching ratios of $T_8\to T_8 P_8$ weak decays. $^\ddagger$denotes which  experimental data have been used to give the effective constraints of the parameters, $^\dagger$denotes the predictions depend on the relative phase, and $^\otimes$denotes  which  experimental data are not used to constrain parameters. }\vspace{0.1cm}
{\footnotesize
\begin{tabular}{lccc}  \hline\hline
 Observables & Exp. Data \cite{PDG2018}& SU(3) IRA  & Isospin Relations  \\\hline
$\mathcal{B}(\Sigma^+\rightarrow p\pi^0)(\times10^{-2})$&$51.57\pm0.30$&$51.57\pm0.30^\ddagger$&$51.57\pm0.30^\ddagger$ \\
$\mathcal{B}(\Sigma^+\rightarrow n\pi^+)(\times10^{-2})$&$48.31\pm0.30$&$48.31\pm0.30^\ddagger$&$48.31\pm0.30^\ddagger$ \\
$\mathcal{B}(\Sigma^-\rightarrow n\pi^-)(\times10^{-2})$&$99.848\pm0.005$&$99.848\pm0.005^\ddagger$&$99.848\pm0.005^\ddagger$ \\
$\mathcal{B}(\Sigma^0\rightarrow p\pi^-)(\times10^{-10})$&$\cdots$&$4.82\pm0.49$&$4.82\pm0.50$ \\
$\mathcal{B}(\Sigma^0\rightarrow n\pi^0)(\times10^{-10})$&$\cdots$&$\cdots$&$2.41\pm0.27$\\\hline
$\mathcal{B}(\Lambda^0\rightarrow p\pi^-)(\times10^{-2})$&$63.9\pm0.5$&$64.19\pm0.21^\ddagger$&$\cdots$  \\
$\mathcal{B}(\Lambda^0\rightarrow n\pi^0)(\times10^{-2})$&$35.8\pm0.5$&$35.42\pm0.12^\ddagger$&$\cdots$ \\\hline
$\mathcal{B}(\Xi^-\rightarrow \Lambda^0\pi^-)(\times10^{-2})$&$99.887\pm0.035$&$99.887\pm0.035^\ddagger$&$\cdots$ \\
$\mathcal{B}(\Xi^0\rightarrow \Lambda^0\pi^0)(\times10^{-2})$&$99.524\pm0.012$&$80.016\pm3.746^\otimes$&$\cdots$ \\\hline
\end{tabular}\label{Tab:T82T8M8Br}}
\end{table}
The experimental measurements with the $\pm1\sigma$ error bar of  $T_8\to T_8 P_8$ weak decays are listed in the second column of Tab. \ref{Tab:T82T8M8Br}.
There are four real parameters ($A_1, A_2e^{i\phi_{A}}, A_3$) for five $\Sigma\rightarrow p\pi,n\pi$ decays,
 one can obtain $A_1=2.48\pm0.01$, $A_2=1.74\pm0.01$ and $|\phi_A|\leq45.35^\circ$ by using the data of $\mathcal{B}(\Sigma^+\rightarrow p\pi^0,n\pi^+,\Sigma^-\rightarrow n\pi^-)$, furthermore, $\mathcal{B}(\Sigma^0\rightarrow p\pi^-)$ could be obtained in terms of $A_1$. In addition, the five $\Sigma\to n\pi,p\pi$ decay modes also have the isospin relations
\begin{eqnarray}
A(\Sigma^+\rightarrow p\pi^0)&=&-\frac{\sqrt{2}}{3}\left(A_{\frac{1}{2}}-A_{\frac{3}{2}}\right),\nonumber\\
A(\Sigma^+\rightarrow n\pi^+)&=&\frac{1}{3}\left(2A_{\frac{1}{2}}+A_{\frac{3}{2}}\right),\nonumber\\
A(\Sigma^-\rightarrow n\pi^-)&=&-\frac{\sqrt{2}}{3}\left(A_{\frac{1}{2}}-A_{\frac{3}{2}}\right),\nonumber\\
A(\Sigma^0\rightarrow p\pi^-)&=&A_{\frac{3}{2}},\nonumber\\
A(\Sigma^0\rightarrow n\pi^0)&=&\frac{1}{3}\left(A_{\frac{1}{2}}+2A_{\frac{3}{2}}\right).\label{Eq:isospin}
\end{eqnarray}
There are three real parameters ($A_{\frac{1}{2}}$, $A_{\frac{3}{2}}e^{i\phi_{13}}$) in Eq.  (\ref{Eq:isospin}). Using the data of $\mathcal{B}(\Sigma^+\rightarrow p\pi^0,\Sigma^+\rightarrow n\pi^+,\Sigma^-\rightarrow n\pi^-)$, one can get $\mathcal{B}(\Sigma^0\rightarrow p\pi^-)$ and $\mathcal{B}(\Sigma^0\rightarrow n\pi^0)$, which are listed in the last column of  Tab. \ref{Tab:T82T8M8Br}. We can see that SU(3) IRA and isospin relations give the consistent predictions for $\mathcal{B}(\Sigma^0\rightarrow p\pi^-)$.

For $\Lambda^0\rightarrow p\pi^-,n\pi^0$ decays, there is only one parameter $A_4$.  We first get the value of $|A_4|$ from the data of  $\mathcal{B}(\Lambda^0\rightarrow p\pi^-)$, then   further considering the experimental data of   $\mathcal{B}(\Lambda^0\rightarrow n\pi^0)$,   finally give the predictions of  $\mathcal{B}(\Lambda^0\rightarrow p\pi^-,n\pi^0)$  in the third column of  Tab. \ref{Tab:T82T8M8Br}.   One can see that the data of both $\mathcal{B}(\Lambda^0\rightarrow p\pi^-)$ and $\mathcal{B}(\Lambda^0\rightarrow n\pi^0)$ give the effective bounds on the parameter $|A_4|$, and  the IRA predictions for $\mathcal{B}(\Lambda^0\rightarrow p\pi^-,n\pi^0)$
are in agreement with the present data.
Noted that, if only considering  the experimental constraint from $\mathcal{B}(\Lambda^0\rightarrow p\pi^-)$,  the prediction of $\mathcal{B}(\Lambda^0\rightarrow p\pi^-)$ given in the third column of Tab. \ref{Tab:T82T8M8Br},  would be completely the same as the experimental datum. The slight difference between the prediction and datum comes from the experimental constraint of $\mathcal{B}(\Lambda^0\rightarrow n\pi^0)$.

For $\Xi^-\rightarrow \Lambda^0\pi^-$ and $\Xi^0\rightarrow \Lambda^0\pi^0$ decays, there is only one parameter $A_5$. We use the data of $\mathcal{B}(\Xi^-\rightarrow \Lambda^0\pi^-)$ to obtain $|A_5|$, and then predict $\mathcal{B}(\Xi^0\rightarrow \Lambda^0\pi^0)$. We obtain $\mathcal{B}(\Xi^0\rightarrow \Lambda^0\pi^0)=(80.016\pm3.746)\%$, which is about $16\%$ smaller than its data.  The reason could be that
 the neglected $C_+$ term or SU(3) breaking effects might give a contribution of a few percent level to
 $\mathcal{B}(\Xi^-\rightarrow \Lambda^0\pi^-)$ and $\mathcal{B}(\Xi^0\rightarrow \Lambda^0\pi^0)$.

\subsubsection{$T_{10}\to T_8M_8$ weak decays}

Feynman diagrams for $T_{10}\to T_{8}M_8$ nonleptonic decays are also displayed in Fig. \ref{fig:NLSU3},
 and the SU(3) IRA amplitudes are
 {\small
\begin{eqnarray}
A(T_{10}\to T_8M_8)^{IRA}&=&
\bar{a}_1H(4)^{lk}_{m}(T_{10})^{nij}(T_{8})_{[ik]j} (M_8)_l^m+\bar{a}_2H(4)^{lk}_{m}(T_{10})^{nij}(T_{8})_{[jk]i} (M_8)_l^m\nonumber\\ &&+\bar{b}_1H(4)^{lk}_{m}(T_{10})^{nij}(T_{8})_{[kl]i}(M_8)_j^m+\bar{b}_2H(4)^{lk}_{m}(T_{10})^{nij}(T_{8})_{[ki]l}(M_8)_j^m +\bar{b}_3H(4)^{lk}_{m}(T_{10})^{nij}(T_{8})_{[il]k} (M_8)_j^m\nonumber\\
&&+\bar{c}_1H(4)^{lk}_{j}(T_{10})^{nij}(T_{8})_{[km]i}(M_8)_l^m
+\bar{c}_2H(4)^{lk}_{j}(T_{10})^{nij}(T_{8})_{[ki]m}(M_8)_l^m+\bar{c}_3H(4)^{lk}_{j}(T_{10})^{nij}(T_{8})_{[im]k} (M_8)_j^m\nonumber\\
&&+\bar{d}_1H(4)^{lk}_{j}(T_{10})^{nij}(T_{8})_{[mk]l} (M_8)_i^m +\bar{d}_2H(4)^{lk}_{j}(T_{10})^{nij}(T_{8})_{[ml]k}(M_8)_i^m+\bar{d}_3H(4)^{lk}_{j}(T_{10})^{nij}(T_{8})_{[kl]m}(M_8)_i^m\nonumber\\
&&+\bar{e}_1H(\bar{2})^{k}(T_{10})^{nij}(T_{8})_{[im]j} (M_8)_k^m+\bar{e}_2H(\bar{2})^{k}(T_{10})^{nij}(T_{8})_{[jm]i} (M_8)_k^m\nonumber\\
&&+\bar{f}_1H(\bar{2})^{k}(T_{10})^{nij}(T_{8})_{[ki]m}(M_8)_j^m+\bar{f}_2H(\bar{2})^{k}(T_{10})^{nij}(T_{8})_{[km]i}(M_8)_j^m+\bar{f}_3H(\bar{2})^{k}(T_{10})^{nij}(T_{8})_{[im]k}(M_8)_j^m.\nonumber\\
\label{Eq:T102T8M8}
\end{eqnarray}}
Considering $H(4)^{lk}_{m}$ and $(T_{10})^{nij}$ is symmetric in upper indices, we have the relations
\begin{eqnarray}
\bar{a}_2=\bar{a}_1, ~~~~~~\bar{b}_1=0, ~~~~~~\bar{b}_3=-\bar{b}_2,~~~~~~\bar{d}_2=\bar{d}_1,~~~~~~\bar{d}_3=0,~~~~~~\bar{e}_2=\bar{e}_1.
\end{eqnarray}
 Then Eq.  (\ref{Eq:T102T8M8}) can be simplified as
 {\small
\begin{eqnarray}
A(T_{10}\to T_8M_8)^{IRA,J}&=&
\bar{a}_1H(4)^{lk}_{m}(T_{10})^{nij}(T_{8})_{[ik]j} (M_8)_l^m+\bar{b}_2H(4)^{lk}_{m}(T_{10})^{nij}(T_{8})_{[ki]l}(M_8)_j^m \nonumber\\
&&+\bar{c}_1H(4)^{lk}_{j}(T_{10})^{nij}(T_{8})_{[km]i}(M_8)_l^m
+\bar{c}_2H(4)^{lk}_{j}(T_{10})^{nij}(T_{8})_{[ki]m}(M_8)_l^m+\bar{c}_3H(4)^{lk}_{j}(T_{10})^{nij}(T_{8})_{[im]k} (M_8)_j^m\nonumber\\
&&+\bar{d}_1H(4)^{lk}_{j}(T_{10})^{nij}(T_{8})_{[mk]l} (M_8)_i^m+\bar{e}_1H(\bar{2})^{k}(T_{10})^{nij}(T_{8})_{[im]j} (M_8)_k^m\nonumber\\
&&+\bar{f}_1H(\bar{2})^{k}(T_{10})^{nij}(T_{8})_{[ki]m}(M_8)_j^m+\bar{f}_2H(\bar{2})^{k}(T_{10})^{nij}(T_{8})_{[km]i}(M_8)_j^m+\bar{f}_3H(\bar{2})^{k}(T_{10})^{nij}(T_{8})_{[im]k}(M_8)_j^m.\nonumber\\
\label{Eq:T102T8M8J}
\end{eqnarray}}
\begin{table}[t]
\renewcommand\arraystretch{1.5}
\tabcolsep 0.08in
\centering
\caption{The  SU(3) IRA amplitudes of $T_{10}\to T_8M_8$ weak decays.}\vspace{0.1cm}
{\footnotesize
\begin{tabular}{rcccc}  \hline\hline
 Amplitudes ~~~~~&$H(4)^{12}_{1}=\frac{1}{3}$&$H(4)^{22}_{2}=-\frac{1}{3}$&$H(\bar{2})^{2}=1$&Simplified Amplitudes  \\\hline
$A(\Omega^-\rightarrow\Xi^0\pi^-)$&$2\bar{a}_1$&&$2\bar{e}_1$ & $\bar{A}_1$\\
$\sqrt{2}A(\Omega^-\rightarrow\Xi^-\pi^0)$&$-2\bar{a}_1$&$2\bar{a}_1$&$2\bar{e}_1$& $\bar{A}_2$ \\
$\sqrt{6}A(\Omega^-\rightarrow\Lambda^0K^-)$&&&$\bar{f}_1+2\bar{f}_2+\bar{f}_3$& $\bar{A}_3$ \\
\hline
$3\sqrt{2}A(\Xi^{*-}\rightarrow\Lambda^0\pi^-)$&$6\bar{a}_1$&$2\bar{c}_1+\bar{c}_2+\bar{c}_3$&$6\bar{e}_1+\bar{f}_1+2\bar{f}_2+\bar{f}_3$ & $3\bar{A}_1+\bar{A}_3$ \\
$\sqrt{6}A(\Xi^{*-}\rightarrow\Sigma^0\pi^-)$&$2\bar{a}_1-4\bar{b}_2$&$\bar{c}_3-\bar{c}_2$&$2\bar{e}_1-\bar{f}_1+\bar{f}_3$& \\
$\sqrt{6}A(\Xi^{*-}\rightarrow\Sigma^-\pi^0)$&$-2\bar{a}_1$&$2\bar{a}_1-2\bar{b}_2+\bar{c}_3-\bar{c}_2$&$2\bar{e}_1-\bar{f}_1+\bar{f}_3$&\\
$\sqrt{3}A(\Xi^{*0}\rightarrow\Sigma^+\pi^-)$&$2\bar{a}_1+\bar{c}_3-\bar{c}_2$&&$2\bar{e}_1$&$\bar{A}_1$\\
$\sqrt{3}A(\Xi^{*0}\rightarrow\Sigma^-\pi^+)$&$\bar{c}_2-\bar{c}_3$&$2\bar{b}_2$&$\bar{f}_1-\bar{f}_3$\\
$6A(\Xi^{*0}\rightarrow\Lambda^0\pi^0)$&$-6\bar{a}_1+2(2\bar{c}_1+\bar{c}_2+\bar{c}_3)$&$6\bar{a}_1$&$6\bar{e}_1+\bar{f}_1+2\bar{f}_2+\bar{f}_3$&$3\bar{A}_2+\bar{A}_3$\\
$2\sqrt{3}A(\Xi^{*0}\rightarrow\Sigma^0\pi^0)$&$2\bar{a}_1-4\bar{b}_2$&$-2\bar{a}_1$&$-2\bar{e}_1-\bar{f}_1+\bar{f}_3$\\
$\sqrt{3}A(\Xi^{*-}\rightarrow nK^-)$&$2\bar{b}_2$&$2\bar{d}_1$&$-\bar{f}_2-\bar{f}_3$\\
\hline
$\sqrt{3}A(\Sigma^{*-}\rightarrow n\pi^-)$&$-2\bar{a}_1+2\bar{b}_2$&$-\bar{c}_1-\bar{c}_3+2\bar{d}_1$&$-2\bar{e}_1-\bar{f}_2-\bar{f}_3$\\
\hline
\end{tabular}\label{Tab:T102T8M8WD}}
\end{table}
The IRA amplitudes for $T_{10}\to T_8P_8$ weak decays are listed in Tab. \ref{Tab:T102T8M8WD}, and the IRA amplitudes for $T_{10}\to T_8V_8$ weak decays have similar relations.
If neglecting  $H(\overline{4})^{22}_{2}$ terms and $c_i$ terms in $H(4)^{12}_1$, and redefining the parameters
\begin{eqnarray}
\bar{A}_1&=&2(\bar{a}_1+\bar{e}_1),\nonumber\\
\bar{A}_2&=&2(-\bar{a}_1+\bar{e}_1),\nonumber\\
\bar{A}_3&=&2(\bar{f}_1+2\bar{f}_2+\bar{f}_2),
\end{eqnarray}
the six decay amplitudes can be given in simpler forms, which are shown in the last column of Tab.\ref{Tab:T102T8M8WD}.
Furthermore, we have the relation $A(\Xi^{*-}\rightarrow\Sigma^0\pi^-)=A(\Xi^{*-}\rightarrow\Sigma^-\pi^0)$ if only
considering the dominant $H(\bar{2})^2$ contributions.

The branching ratios of $T_{10}\to T_8 P_8$ can be obtained in terms of IRA amplitudes
\begin{eqnarray}
\mathcal{B}(T_{10A}\to T_{8B} P_8)=\frac{\tau_A|p_{cm}|}{16\pi m_A^2}\big|A(T_{10A}\to T_{8B} P_8)\big|^2,
\end{eqnarray}
and the mass difference in $A(T_{10A}\to T_{8B} P_8)$, which is similar to Eq.  (\ref{Eq:AMDT8}), is also considered.

At present, only three $\Omega^-$ decay modes have been measured
\begin{eqnarray}
\mathcal{B}(\Omega^-\rightarrow\Xi^0\pi^-)(\times10^{-2})&=&(23.6\pm0.7)\times10^{-2},\nonumber\\
\mathcal{B}(\Omega^-\rightarrow\Xi^-\pi^0)(\times10^{-2})&=&(8.6\pm0.4)\times10^{-2},\nonumber\\
\mathcal{B}(\Omega^-\rightarrow\Lambda^0K^-)(\times10^{-2})&=&(67.8\pm0.7)\times10^{-2}.
\end{eqnarray}
We obtain that $|\bar{A}_1|=8.54\pm0.19$, $|\bar{A}_2|=7.47\pm0.23$ and $|\bar{A}_3|=5.36\pm0.08$   from the data of $\mathcal{B}(\Omega^-\rightarrow\Xi^0\pi^-)$, $\mathcal{B}(\Omega^-\rightarrow\Xi^-\pi^0)$ and $\mathcal{B}(\Omega^-\rightarrow\Lambda^0K^-)$, respectively.
Then we predict that
\begin{eqnarray}
\mathcal{B}(\Xi^{*-}\rightarrow\Lambda^0\pi^-)&=&(1.06\pm0.90)\times10^{-12},\nonumber\\
\mathcal{B}(\Xi^{*0}\rightarrow\Sigma^+\pi^-)&=&(5.96\pm0.58)\times10^{-14},\nonumber\\
\mathcal{B}(\Xi^{*0}\rightarrow\Lambda^0\pi^0)&=&(5.02\pm4.06)\times10^{-13},
\end{eqnarray}
where the prediction of $\mathcal{B}(\Xi^{*-}\rightarrow\Lambda^0\pi^-)$ depends  on the relative phase between $\bar{A}_1$  and $\bar{A}_3$, and the prediction of $\mathcal{B}(\Xi^{*0}\rightarrow\Lambda^0\pi^0)$ depends  on the relative phase  between $\bar{A}_2$  and $\bar{A}_3$.

\subsection{Electromagnetic or strong decays of light baryons }
\begin{figure}[ht]
\centering
\includegraphics[scale=0.5]{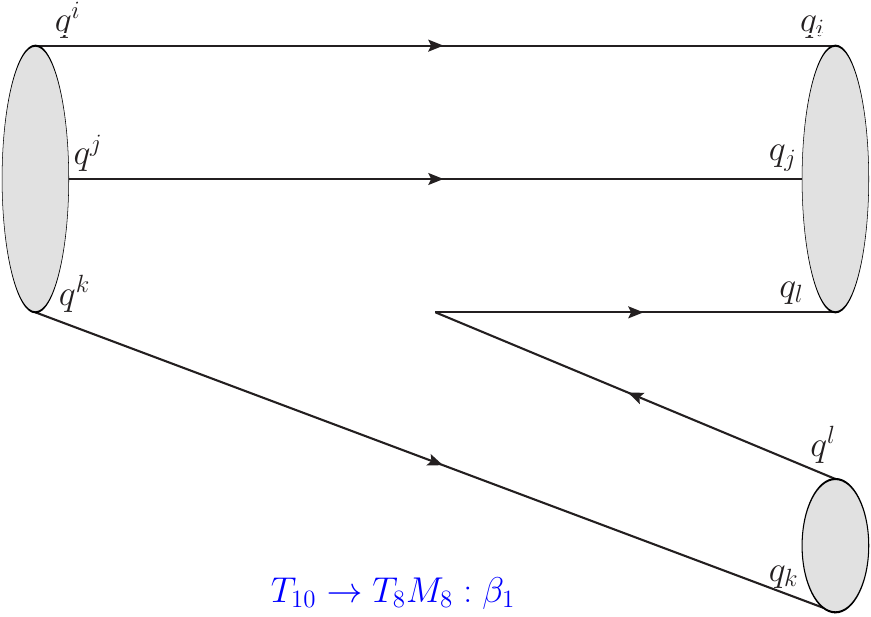}
\caption{Feynman diagram of IRA  for $T_{10}\to T_{8}M_8$  ES decays. }
\label{fig:NLSU3ES}
\end{figure}
The light baryons $T_{10}$ can also decay through  electromagnetic or strong interactions.  The Feynman diagram of electromagnetic or strong (ES) decays
of $T_{10}$ is shown in Fig. \ref{fig:NLSU3ES}. In this case, we only need consider the SU(3) symmetry between initial and final states.
The SU(3) IRA amplitude of  $T_{10}\to T_{8}M_8$ ES decay  is
\begin{eqnarray}
A(T_{10}\to T_8M_8)^{ES,IRA}= \beta_1(T_{10})^{ijk}(T_{8})_{[il]j}(M_8)^l_k.
\label{Eq:T102T8M8d}
\end{eqnarray}
There is only one parameter $\beta_1$ for these IRA amplitude. The IRA amplitudes of all the ES $T_{10}\to T_8P_8$ decays are given
 in  Tab. \ref{Tab:T102T8M8AESD}.

\begin{table}[t]
\renewcommand\arraystretch{1.5}
\tabcolsep 0.6in
\centering
\caption{The IRA amplitudes of $T_{10}\to T_8P_8$  ES decays  under the SU(3) flavor symmetry.}\vspace{0.1cm}
{\footnotesize
\begin{tabular}[t]{rc}  \hline\hline
Amplitudes~~~~~~& SU(3) IRA amplitudes\\\hline
$\sqrt{6}A(\Sigma^{*+}\rightarrow\Sigma^0\pi^+)$&$\beta_1$ \\
$\sqrt{6}A(\Sigma^{*+}\rightarrow\Sigma^+\pi^0)$&$\beta_1$ \\
$2\sqrt{6}A(\Sigma^{*0}\rightarrow\Sigma^0\pi^0)$&$0$ \\
$\sqrt{6}A(\Sigma^{*0}\rightarrow\Sigma^+\pi^-)$&$\beta_1$\\
$\sqrt{6}A(\Sigma^{*0}\rightarrow\Sigma^-\pi^+)$&$-\beta_1$ \\
$\sqrt{6}A(\Sigma^{*-}\rightarrow\Sigma^-\pi^0)$&$\beta_1$\\
$\sqrt{6}A(\Sigma^{*-}\rightarrow\Sigma^0\pi^-)$&$\beta_1$\\\hline
$3\sqrt{2}A(\Sigma^{*+}\rightarrow\Lambda^0\pi^+)$&$-3\beta_1$ \\
$6\sqrt{2}A(\Sigma^{*0}\rightarrow\Lambda^0\pi^0~)$&$6\beta_1$\\
$3\sqrt{2}A(\Sigma^{*-}\rightarrow\Lambda^0\pi^-)$&$3\beta_1$ \\\hline
$\sqrt{6}A(\Xi^{*0}\rightarrow\Xi^0\pi^0~)$&$\beta_1$\\
$\sqrt{3}A(\Xi^{*0}\rightarrow\Xi^-\pi^+)$&$-\beta_1$ \\
$\sqrt{6}A(\Xi^{*-}\rightarrow\Xi^-\pi^0~)$&$\beta_1$ \\
$\sqrt{3}A(\Xi^{*-}\rightarrow\Xi^0\pi^-)$&$\beta_1$ \\\hline
%
%
%
\end{tabular}\label{Tab:T102T8M8AESD}}
\end{table}
\begin{table}[t]
\renewcommand\arraystretch{1.4}
\tabcolsep 0.25in
\centering
\caption{Branching ratios of $T_{10}\to T_8P_8$ ES decays within $1\sigma$ error. $^\ddagger$denotes which  experimental data have been used to give the effective constraints on the parameters, and $^\otimes$denotes which experimental data are not used to constrain parameters. }\vspace{0.1cm}
{\footnotesize
\begin{tabular}[t]{lcc}  \hline\hline
 Branching ratios &Exp.&SU(3) IRA \\\hline
$\mathcal{B}(\Sigma^{*+}\rightarrow\Sigma^0\pi^+)(\times10^{-2})$&$\cdots$&$5.34\pm0.50$ \\
$\mathcal{B}(\Sigma^{*+}\rightarrow\Sigma^+\pi^0)(\times10^{-2})$&$\cdots$&$6.59\pm0.61$ \\
$\mathcal{B}(\Sigma^{*0}\rightarrow\Sigma^0\pi^0)(\times10^{-2})$&$\cdots$&$0$\\
$\mathcal{B}(\Sigma^{*0}\rightarrow\Sigma^+\pi^-)(\times10^{-2})$&$\cdots$&$6.20\pm0.78$\\
$\mathcal{B}(\Sigma^{*0}\rightarrow\Sigma^-\pi^+)(\times10^{-2})$&$\cdots$&$4.71\pm0.59$ \\
$\mathcal{B}(\Sigma^{*-}\rightarrow\Sigma^-\pi^0)(\times10^{-2})$&$\cdots$&$5.40\pm0.60$ \\
$\mathcal{B}(\Sigma^{*-}\rightarrow\Sigma^0\pi^-)(\times10^{-2})$&$\cdots$&$5.66\pm0.63$\\
$\mathcal{B}(\Sigma^{*}\rightarrow\Sigma\pi)(\times10^{-2})$&$11.7\pm1.5$&$11.24\pm0.28$ \\\hline
$\mathcal{B}(\Sigma^{*+}\rightarrow\Lambda^0\pi^+)(\times10^{-2})$&$\cdots$&$86.14\pm7.62$ \\
$\mathcal{B}(\Sigma^{*0}\rightarrow\Lambda^0\pi^0)(\times10^{-2})$&$\cdots$&$91.68\pm11.36$\\
$\mathcal{B}(\Sigma^{*-}\rightarrow\Lambda^0\pi^-)(\times10^{-2})$&$\cdots$&$84.44\pm8.96$\\
$\mathcal{B}(\Sigma^{*}\rightarrow\Lambda^0\pi)(\times10^{-2})$&$87.0\pm1.5$&$87.00\pm1.50^\ddag$ \\\hline
$\mathcal{B}(\Xi^{*0}\rightarrow\Xi^0\pi^0)(\times10^{-2})$&$\cdots$&$48.22\pm6.55$\\
$\mathcal{B}(\Xi^{*0}\rightarrow\Xi^-\pi^+)(\times10^{-2})$&$\cdots$&$76.23\pm10.32$ \\
$\mathcal{B}(\Xi^{*-}\rightarrow\Xi^-\pi^0)(\times10^{-2})$&$\cdots$&$43.05\pm11.01$ \\
$\mathcal{B}(\Xi^{*-}\rightarrow\Xi^0\pi^-)(\times10^{-2})$&$\cdots$&$94.33\pm24.12$ \\
$\mathcal{B}(\Xi^{*}\rightarrow\Xi\pi)(\times10^{-2})$&$100$&$131.01\pm24.40^\otimes$ \\\hline
\end{tabular}\label{Tab:T102T8M8duudNR}}
\end{table}
For these ES decays, only three branching ratios are measured, which are given in Tab.  \ref{Tab:T102T8M8duudNR}.
 We first get $|\beta_1|$ from the data of $\mathcal{B}(\Sigma^{*}\rightarrow\Sigma\pi)$, then also consider  the experimental constraint from  $\mathcal{B}(\Sigma^{*}\rightarrow\Lambda\pi)$, and
 finally give the predictions of other specific branching ratios.
 Our SU(3) IRA predictions  are given in  Tab. \ref{Tab:T102T8M8duudNR}, where
one can see that, within $1\sigma$ error, the experimental result of $\mathcal{B}(\Sigma^{*}\rightarrow\Lambda\pi)$ can effectively constrain $|\beta_1|$. In addition,  when IRA predictions are consistent with the data of $\mathcal{B}(\Sigma^{*}\rightarrow\Sigma\pi)$ and $\mathcal{B}(\Sigma^{*}\rightarrow\Lambda\pi)$, the prediction of $\mathcal{B}(\Xi^{*}\rightarrow\Xi\pi)$  is slightly larger than its experimental result,  which might imply that the SU(3) breaking effects could give visible contributions to $\mathcal{B}(\Xi^{*}\rightarrow\Xi\pi)$.  Nevertheless, the prediction and experimental data of $\mathcal{B}(\Xi^{*}\rightarrow\Xi\pi)$ can be consistent within $1.3\sigma$ error.
 And moreover, the decay width predictions of $\Xi^{*0}\rightarrow\Xi\pi$ and $\Xi^{*-}\rightarrow\Xi\pi$ in the chiral quark-soliton
model  are also slightly larger than their experimental data \cite{Yang:2018idi}.

Note that the ES $T_{8}\to T_8 P_8$ decays and the ES $T_{10}\to T_8 K$ decays are not allowed by the phase space, since the sum of final hadron masses is larger than the mass of initial state.

\section{SUMMARY}
Light baryon decays play very important role in testing the SM and searching for new physics beyond the SM. Many decay modes have been measured and some decays can be studied at BESIII and LHCb experiments now. Motivated by this, we have analyzed the semileptonic decays and two-body nonleptonic decays of light baryon octet and decuplet
by using the irreducible representation approach to test the SU(3) flavor symmetry. Our main results can be summarized as follows:
\begin{itemize}
\item{\bf Semileptonic light baryon decays}: We find that all  branching ratio predictions  of octet and decuplet baryons through $s\to u \ell^-\bar{\nu}_\ell$ and $d\to u e^-\bar{\nu}_e$  transitions  with SU(3) IRA in $S_2$ case are quite consistent with present experimental measurements within $1\sigma$ error.
We predict that $\mathcal{B}(\Xi^{-}\rightarrow \Sigma^0\mu^-\bar{\nu}_\mu)$ and $\mathcal{B}(\Omega^-\rightarrow\Xi^0\mu^-\bar{\nu}_\mu)$
are at the order of magnitudes of $10^{-6}$ and $10^{-3}$, respectively, and $\mathcal{B}(\Sigma^-\rightarrow \Sigma^0e^-\bar{\nu}_e,
\Xi^-\rightarrow \Xi^0e^-\bar{\nu}_e)$ are at the order of $10^{-10}$. These decays are promising to be observed by the BESIII and LHCb
experiments or the future experiments. However, other branching ratios, which are in the range of $10^{-20}-10^{-13}$,
may not be measured for a long time. Moreover, the longitudinal branching ratios of decays of $T_{8A} \to T_{8B} \ell^- \bar{\nu}_\ell$ are also
predicted in this work.

\item{\bf Nonleptonic two-body light baryon decays}: We obtain  the relations of different decay amplitudes by the SU(3) IRA and isospin symmetry. In $T_8\to T_8 P_8$ weak decays, we find that SU(3) IRA predictions of the branching ratios of $\Sigma,\Lambda$ baryons are consistent with present experimental data,  $\mathcal{B}(\Sigma^0\to p\pi^-,n\pi^0)$ are at the order of $10^{-10}$ by  the SU(3) IRA or isospin symmetry, and the neglected $C_+$ terms or SU(3) symmetry breaking effects might give a contribution of a few percent
 to the two branching ratios of $\Xi\to\Lambda\pi$. In $T_{10}\to T_8 P$ weak decays, we predict that  $\mathcal{B}(\Xi^{*-}\rightarrow\Lambda^0\pi^-)$,  $\mathcal{B}(\Xi^{*0}\rightarrow\Lambda^0\pi^0)$ and  $\mathcal{B}(\Xi^{*0}\rightarrow\Sigma^+\pi^-)$ are at the orders of $10^{-12}$, $10^{-13}$ and $10^{-14}$, respectively. In $T_{10}\to T_8 P_8$ ES decays,  when IRA predictions are consistent with the data of $\mathcal{B}(\Sigma^{*}\rightarrow\Sigma\pi)$ and $\mathcal{B}(\Sigma^{*}\rightarrow\Lambda\pi)$, the prediction of $\mathcal{B}(\Xi^{*}\rightarrow\Xi\pi)$  is slightly larger than experimental data,  which imply that the SU(3) symmetry breaking effects could give visible contributions to $\mathcal{B}(\Xi^{*}\rightarrow\Xi\pi)$. In addition, we  given all the specific branching ratio predictions for these $T_{10}\to T_8 P_8$ ES decays.
\end{itemize}

Although flavor SU(3) symmetry is approximate, it can still provide us very useful
information about these decays. According to our predictions, some branching ratios are accessible to the experiments at BESIII and LHCb. Our results in this work can be used to test SU(3) flavor symmetry approach in light baryon decays by the future experiments..

\section*{ACKNOWLEDGEMENTS}
Ru-Min Wang thanks  Wei Wang for helpful communications.
The work was supported by the National Natural Science Foundation of China (Contract Nos. 11675137, 11875168 and 11875054),
the Joint Large-Scale Scientific Facility Funds of the NSFC and CAS under Contract No. U1532257, CAS under Contract No. QYZDJ-SSWSLH003; and
and the National Key Basic Research Program of China under Contract No. 2015CB856700.

\section*{References}
\renewcommand{\baselinestretch}{1.43}

\end{document}